\documentclass{emulateapj}
\usepackage{amsmath,amssymb,graphicx}

\newcommand{\Ob}{\Omega_b}
\newcommand{\Om}{\Omega_m}
\newcommand{\Ol}{\Omega_\Lambda}
\newcommand{\Msun}{M_\odot}
\newcommand{\Zsolar}{Z_\odot}
\newcommand{\sfr}{\dot{\rho}_*}
\newcommand{\taue}{\tau_{\rm e}}

\newcommand{\HI}{{\rm HI}}
\newcommand{\HII}{{\rm HII}}

\newcommand{\HeI}{{\rm HeI}}
\newcommand{\HeII}{{\rm HeII}}
\newcommand{\HeIII}{{\rm HeIII}}
\newcommand{\nH}{n_{\rm H}}

\newcommand{\nHII}{n_\HII}
\newcommand{\nelec}{n_{\rm e}}
\newcommand{\fHI}{{\nu{\rm HI}}}
\newcommand{\fHeI}{{\nu{\rm HeI}}}
\newcommand{\fHeII}{{\nu{\rm HeII}}}
\newcommand{\ngam}{n_\gamma}
\newcommand{\gHI}{{\gamma{\rm HI}}}

\newcommand{\lp}{l_{\rm p}}
\newcommand{\Ap}{A_{\rm p}}
\newcommand{\Vp}{V_{\rm p}}

\newcommand{\Mvir}{M_{\rm vir}}
\newcommand{\Rvir}{R_{\rm vir}}
\newcommand{\Tvir}{T_{\rm vir}}
\newcommand{\tdyn}{t_{\rm dyn}}

\newcommand{\PopIII}{{\rm PopIII}}

\newcommand{\ipix}{i_{\rm pix}}
\newcommand{\lbase}{l_{\rm base}}
\newcommand{\lmerge}{l_{\rm merge}}
\newcommand{\Nbase}{N_{\rm base}}
\newcommand{\Nmerge}{N_{\rm merge}}
\newcommand{\Nrt}{N_{\rm RT}}
\newcommand{\Nray}{N_{\rm ray}}
\newcommand{\Nrayrh}{N_{\rm ray/RT}}
\newcommand{\Nraymax}{N_{\rm ray,max}}
\newcommand{\Nrad}{N_{\rm rad}}

\newcommand{\deriv}{{\rm d}}
\newcommand{\dt}{{\rm d}t}
\newcommand{\dz}{{\rm d}z}
\newcommand{\dnu}{{\rm d}\nu}
\newcommand{\ts}{\Delta t}
\newcommand{\bx}{\boldsymbol{x}}
\newcommand{\bvel}{\boldsymbol{v}}

\shorttitle{Radiative transfer simulations of cosmic reionization}
\shortauthors{Trac \& Cen}

\begin{document}

\title{Radiative Transfer Simulations of Cosmic Reionization I:  Methodology and Initial Results}

\author{Hy Trac and Renyue Cen}
\affil{Department of Astrophysical Sciences, Princeton University, Princeton, New Jersey, 08544, USA}
\email{htrac@astro.princeton.edu}
\email{cen@astro.princeton.edu}

\begin{abstract}
We present a new hybrid code for large volume, high resolution simulations of cosmic reionization, which utilizes a N-body algorithm for dark matter, physically motivated prescriptions for baryons and star formation, and an adaptive ray tracing algorithm for radiative transfer of ionizing photons.  Two test simulations each with 3 billion particles and 400 million rays in a 50 Mpc/$h$ box have been run to give initial results.  Halos are resolved down to virial temperatures of $10^4$ K for the redshift range of interest in order to robustly model star formation and clumping factors.  This is essential to correctly account for ionization and recombination processes.  We find that the halos and sources are strongly biased with respect to the underlying dark matter, re-enforcing the requirement of large simulation boxes to minimize cosmic variance and to obtain a qualitatively correct picture of reionization.  We model the stellar initial mass function (IMF), by following the spatially dependent gas metallicity evolution, and distinguish between the first generation, Population III (PopIII) stars and the second generation, Population II (PopII) stars.  The PopIII stars with a top-heavy IMF produce an order of magnitude more ionizing photons at high redshifts $z\gtrsim10$, resulting in a more extended reionization.  In our simulations, complete overlap of HII regions occurs at $z\approx6.5$ and the computed mass and volume weighted residual HI fractions at $5\lesssim z\lesssim 6.5$ are both in good agreement with high redshift quasar absorption measurements from SDSS.  The values for the Thomson optical depth are consistent within $1-\sigma$ of the current best-fit value from third-year WMAP.
\end{abstract}


\section{Introduction}

Cosmic reionization starts when the first generation of stars begin to photoionize beyond the interstellar medium (ISM) and into the intergalactic medium (IGM).  The overall picture is complex, with constant creation and possible demise, as well as mergers of individual HII regions.  The mean ionized fraction will grow as the cosmic star formation rate increases.  When a complete overlap of HII regions occurs, reionization comes to an end and the universe becomes transparent to UV photons.  For recent reviews, see \citet{2001BarkanaLoeb} and \citet{2006FCK}.

Currently, there are two major observational constraints on the reionization epoch.  SDSS observations indicate that reionization ends at $z\sim 6$ because spectra of quasars show lack of complete Gunn-Peterson absorption at lower redshifts, while complete Lyman alpha absorption is found at immediately higher redshifts \citep[e.g.][]{2001Fan, 2001Becker, 2006Fan}.  WMAP polarization measurements have yielded a Thomson optical depth of $\tau=0.09\pm0.03$, which suggests that the IGM was largely ionized by redshift $z\sim10$ \citep{2006PageWMAP3, 2007SpergelWMAP3}.  The combination of these two observations imply that reionization may be considerably more complex than the simple assumption of an impulsive phase transition.  Current and future observations, primarily high redshift surveys of Lyman alpha emitting galaxies \citep[e.g.][]{2006Kashikawa}, CMB measurements  of the kinetic Sunyaev-Zeldovich (KSZ) effect from free electrons (e.g.~ACT, SPT, Planck), and radio observations of the 21 cm radiation from neutral hydrogen (e.g.~LOFAR, MWA, SKA), will provide significantly more detailed information.

Progress on the theoretical front have mostly come from analytical and semi-analytical modelling \citep[e.g.][]{2001BarkanaLoeb, 2003Cena, 2003Cenb, 2003WyitheLoeb, 2004BarkanaLoeb, 2004FZH, 2004Madau, 2004RicottiOstriker, 2006WyitheCen}.  Analytical models, based on simplified approaches, are able to afford continuous dynamic range and economically provide insightful information.  However, they are generally limited in scope and applicability.  The semi-analytical approach provides relatively more realistic treatment of the formation of structure and HII regions.  In the most recent work to date \citep[e.g.][]{2007Zahn, 2007Mesinger}, realistic halo distributions and ionization maps have been generated by combining the excursion-set formalism with Lagrangian perturbation theory.  These algorithms are computationally fast and are very useful for exploring the parameter space, especially when the prescriptions are calibrated in detail with realistic simulations.

Numerical simulations directly solve the nonlinear physics of gravitational collapse, hydrodynamics, and radiative transfer.  Furthermore, simulations allow a more straightforward and robust implementation of intricate prescriptions for astrophysical processes, such as star formation, which currently are not simulated from first principles.  However, simulations of reionization are costly and hence, restricted in dynamic range because of limitations in numerical algorithms and computational resources.  High resolution but small volume simulations \citep[e.g.][]{2000Gnedin, 2002Razoumov} can resolve dwarf galaxies, Lyman limit systems, and mini-halos, allowing for more accurate determination of the star formation rates, photoionization rates, and clumping factors.  However, they give highly biased results due to cosmic variance.  Large volume but low resolution simulations \citep[e.g.][]{2003Sokasian, 2003Ciardi, 2005KGH, 2006Iliev, 2006IMSP, 2007Zahn, 2007McQuinn} enable the study of the large-scale structure of the dark matter, baryons, sources, and HII regions, but they must correct for relevant processes happening on unresolved mass, spatial, and temporal scales.

Until now, no other work reported in the literature has simultaneously satisfied the following two demands.  On one hand, high resolution is required to resolve high redshift halos with masses $\sim10^{7.5}-10^{8.5}\Msun/h$, where the majority of photoionizing sources are thought to reside.  Probing small scales is also essential to correctly calculate clumping factors and recombination rates.  On the other hand, it has been argued that a large simulation volume of $\sim(100$ Mpc$/h)^3$ is necessary in order to fairly sample highly biased sources \citep{2004BarkanaLoeb} and large HII regions with characteristic sizes $\sim10-50$ Mpc/$h$ \citep{2004FZH}.  The demand of having at least 10 particles to identify a halo in a simulation with a volume of $(100$ Mpc$/h)^3$ translates to a requirement of more than 20 billion particles.  In this paper, we describe the methodology developed to cross this threshold of mass dynamic range and will present simulations with up to 24 billion particles in companion papers \citep[Trac et al.~2007]{2007Shin}.

Three-dimensional radiative transfer (RT) calculations in cosmological simulations are very challenging because the radiation field can be quite complex.  By ray tracing, one can accurately propagate the radiation field, but it is computationally very expensive.  Since the total number of rays emitted by a source scales with the size of the simulation $N$ and the number of sources also scales with $N$, too many rays are generated with this ${\cal O}(N^2)$ scaling.  The prefactor is also relatively large because the number of sources can be as large as $\sim 10\%$ of $N$.  We have developed an accurate and efficient adaptive ray tracing algorithm where we use the ray splitting scheme of \citet{2002AbelWandelt} to obtain excellent angular resolution for ray tracing from point sources, but have introduced a new ray merging scheme in order to avoid the ${\cal O}(N^2)$ scaling problem.  With adaptive merging of co-parallel rays, the algorithm converges to ${\cal O}(N)$ scaling as the radiation filling factor approaches unity.

In this paper, we use the cosmological parameters:  $\Om=0.26$, $\Ol=0.74$, $\Ob=0.044$, $h=0.72$, $\sigma_8=0.77$, and $n_s=0.95$, based on the latest results from WMAP, SDSS, BAO, SN, and HST \citep[see][and references therein]{2007SpergelWMAP3}.  \S\ref{sec:methods} describes the numerical methods:  a N-body particle-multi-mesh \citep[PMM;][]{2006TracPenOCH} code for the gravitational evolution of the dark matter, a hydrostatic equilibrium model for gas in halos, a star formation prescription for gas cooling in halos, and a RT algorithm with adaptive ray tracing.  \S\ref{sec:simulations} presents two high-resolution simulations each with 3 billion particles, 400 million rays, and $180^3$ RT grid cells in 50 Mpc$/h$ boxes.  In one simulation, we consider only Population II (PopII) stars with a Salpeter initial mass function (IMF), while in the second simulation, we include Population III (PopIII) stars with a top-heavy IMF.  \S\ref{sec:results} discusses initial results: halo mass functions, halo clustering, star formation rates, ionization fractions, clumping factors, and optical depths.  We demonstrate that the current observational  constraints require a star formation history where both PopIII and PopII stars are significant sources of photoionizing radiation.

\section{Numerical Methods}
\label{sec:methods}

\subsection{Dark matter}

We use a particle-multi-mesh \citep[PMM;][]{2006TracPenOCH} N-body code, based on an improved particle-mesh (PM) algorithm, to compute the gravitational dynamics of the collisionless dark matter.  In principle, standard PM codes can achieve high spatial resolution with a large mesh but this comes at a great cost in memory and to a lesser extent in work.  In practice, they are normally limited to a mean interparticle spacing to mesh cell spacing ratio of 2:1, where the storage requirements for particles and grids are approximately balanced.  PMM utilizes a domain-decomposed, FFT-based gravity solver to achieve higher spatial resolution without increasing memory costs.  It is based on a two-level mesh Poisson solver where the gravitational forces are separated into long-range and short-range components.  The long-range force is computed on the root-level, global mesh, much like in a PM code.  To achieve higher spatial resolution, the domain is decomposed into cubical regions and the short-range force is computed on a refinement-level, local mesh.  In the current version, PMM can achieve a spatial resolution of 4 times better than a standard PM code at the same cost in memory.  The simulations in this paper were run with a mean interparticle spacing to mesh cell spacing ratio of 4:1.

For each N-body particle of fixed mass, we store its comoving position $\bx$ and velocity $\bvel$ in order to solve the equations of motion in an expanding Friedman-Robertson-Walker (FRW) universe.  In addition, the local matter density $\rho_m$ associated with each particle is calculated using a SPH-like smoothing kernel.  These seven particle variables are updated every PM time step of $\ts_{PM}\lesssim10^7$ years.  

We locate dark matter halos using a friends-of-friends algorithm with a standard linking length of 0.2 times the mean interparticle spacing.  Halo catalogs are produced containing the following information:  redshift, comoving positions and velocities, and mass.  In addition, we tag each particle that is bound to the identified halos in order to model baryons and star formation.

In a $SCDM$ cosmology with $\Om=1$ and $\Ol=0$, the spherical top-hat collapse model \citep{1980Peebles} defines a virialized halo to be a sphere of mass $\Mvir$ and radius $\Rvir$ with a characteristic average density equal to $\Delta_c=18\pi^2$ times the critical density $\rho_{crit}(z)$ \citep{1993LaceyCole, 1994LaceyCole}.  In a $\Lambda$CDM cosmology with $\Ol\neq0$, the characteristic average density for virialization has been calculated using numerical simulations \citep{1998BryanNorman}, but note that at the high redshifts $z\gtrsim6$ prior to complete reionization, the universe is basically Einstein-de-Sitter and $\Delta_c\approx18\pi^2$.

\subsection{Baryons}

On large, linear scales, we assume the baryons are unbiased tracers of the dark matter because the hydrodynamics of the baryons is primarily dictated by the large-scale gravitational dynamics of the matter and not by the thermodynamics of the gas.  On small, nonlinear scales, the collisionless dark matter continues to collapse, undergoing violent relaxation and forming virialized halos with deep potential wells.  The collisional baryons fall into the potential wells, get shock heated to approximately the virial temperature, and the collapse gets halted by gas pressure.  However, this collisional component can dissipate energy through radiative cooling and eventually collapse to very high densities to form stars.

We approximate the local baryon density $\rho_b$ associated with each particle by multiplying the matter density $\rho_m$ by the cosmic baryon fraction $\Ob/\Om$, except in high density regions where gravitational collapse and pressure support are important.  In virialized halos, we calculate the gas density profile assuming the gas is in hydrostatic equilibrium with the gravitational potential.

For a virialized halo with mass $\Mvir$ and radius $\Rvir$, the dark matter density profile is taken to follow the NFW  formula \citep{1997NFW}:
\begin{equation}
\rho_{dm}(x) = \frac{\Mvir c^3/(4\pi\Rvir^3)}{\ln(1+c)-c/(1+c)}\frac{1}{x(1+x)^2} ,
\end{equation}
where $x\equiv r/r_s$, $r_s$ is a scale radius, and $c\equiv\Rvir/r_s$ is a concentration parameter.  Simulations of early structure formation in a $\Lambda$CDM cosmology produce dark matter halos which are well-fitted by the NFW formula for redshifts $z\lesssim12$ and with only small deviations for redshifts up to $z\sim49$ \citep{2005Gao}.  The concentration parameter is approximately given by \citep[see][and references therein]{2004Dolag},
\begin{equation}
c(\Mvir,z) = \frac{7.5}{1+z}\left(\frac{\Mvir}{M_*}\right)^{-0.1} ,
\label{eqn:concentration}
\end{equation}
where the nonlinear mass is $M_*=2.09\times10^{12}\Msun/h$ for the chosen cosmology.  The high resolution of the simulations is sufficient to locate halos and measure their masses, but it is not enough to accurately measure their dark matter profiles.  Therefore, the NFW profile and the  parametric model for the concentration parameter are used instead.

Assuming hydrostatic equilibrium and a polytropic equation of state $P_g\propto\rho_g^\gamma$, the gas density and temperature profiles can be parametrized as,
\begin{align}
\rho_g(x) 	& = \rho_0 y_g(x)\ ,\\
T_g(x) 	& = T_0 y_g^{\gamma-1}(x)\ ,
\end{align}
where $\gamma$ is the polytropic index and the coefficients $\rho_0\equiv\rho_g(0)$ and $T_0\equiv T_g(0)$ are two boundary conditions.  The dimensionless gas density profile has the analytical solution \citep[e.g.][]{2001KomatsuSeljak},
\begin{equation}
y_g(x) = \left\{1-A\left[1-\frac{\ln(1+x)}{x}\right] \right\}^{1/(\gamma-1)} ,
\end{equation}
where
\begin{equation}
A = \left(\frac{\gamma-1}{\gamma}\right)\left(\frac{G\Mvir\mu}{kT_0\Rvir}\right)\left[\frac{\ln (1+c)}{c}-\frac{1}{1+c}\right]^{-1}\ .
\end{equation}
and $\mu$ is the mean molecular weight.  

We have chosen $\gamma=1.2$ as suggested by numerical simulations and semi-analytic models \citep[see][and references therein]{2005OstrikerBB}.  The baryon fraction at the virial radius is normalized to be equal to the cosmic value.  We also normalize the temperature at the virial radius to be equal to the halo virial temperature,
\begin{equation}
\Tvir = \frac{G\Mvir\mu}{3k\Rvir} = \frac{\mu}{3k}\left(\frac{\Delta_c}{2}\right)^{1/3}[G\Mvir H(z)]^{2/3} .
\end{equation}
Given a dark matter density, we first analytically solve a cubic equation to find the radius and then the corresponding baryon density.  This is done for every particle identified as belonging to a halo and for every PM timestep.

\subsection{Star formation}

\begin{figure}[t]
\center
\includegraphics[width=3.4in]{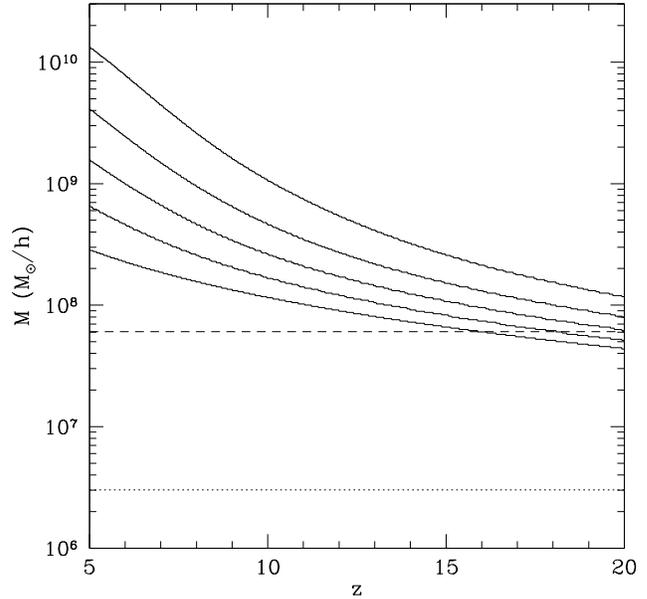}
\caption{The mass range of halos where star formation can take place.  The curve $M(z,f)$ gives the mass range   $M \geq M(z,f)$ which accounts for the fraction $f$ of the total star formation at redshift $z$.  The five solid curves from top to bottom are for the fractions $f$ = (0.2, 0.4, 0.6, 0.8, 1.0).  For comparison, the simulations all have a particle mass of $m_{\rm p}=3.02\times10^6\Msun/h$ (dotted line) and a minimum halo mass of $M_{\rm min} = 20m_{\rm p}$ (dashed line).}
\label{fig:msf}
\end{figure}

Halos with virial temperatures $\gtrsim 10^4$ K are thought to form stars more efficiently than mini-halos.  Above this transition temperature, efficient atomic line cooling allows the gas to dissipate energy and collapse to high enough densities for molecular hydrogen (H2) formation.  Molecular cooling then allows the gas to further collapse to the very high densities needed for star formation in giant molecular clouds.  Star formation in mini-halos is possible because of H2 formation and molecular cooling in dense self-shielded gas.  However, the H2 in mini-halos are very susceptible to dissociation by Lyman-Werner photons from the first stars \citep{1997HRL}.  Therefore, we will assume that mini-halos make only minor contributions to the total stellar density and we model star formation only in halos which cool through atomic transitions.

In our simulations, the star formation criteria resemble those used in hydrodynamic simulations \citep[e.g.][]{1992CenOstriker}.  The first criterion for star formation is that it only occurs in halos with virial temperatures above $10^4$ K.  The second criterion is that the local cooling time must be shorter than the local dynamical time:
\begin{align}
t_{\rm cool} 	& = \frac{3\rho_gkT}{2\mu}\left[n_H^2(\Lambda-\Gamma)\right]^{-1} ,\\
\tdyn			& = \sqrt{\frac{3\pi}{32G\rho_m}} .
\end{align}
where $X$ is the hydrogen fraction by mass and $n_H^2(\Lambda-\Gamma)$ is the net cooling rate per unit volume \citep[see][]{1992Cen, 1998Theuns}.  We calculate the cooling radius as a function of virial mass and redshift assuming that the gas in the halos are in ionization equilibrium, which is a good approximation because of the high densities.  Feedback from photoionization and supernova will affect the cooling radius, but modeling these affects is beyond the scope of this paper and will be left for future work.  Figure \ref{fig:msf} shows the mass range of halos where star formation can take place.  The curve $M(z,f)$ gives the mass range $M \geq M(z,f)$ which accounts for the fraction $f$ of the total star formation at redshift $z$.  The five curves from top to bottom are for the fraction $f$ = (0.2, 0.4, 0.6, 0.8, 1.0) of the total star formation rate.

For every particle within the cooling radius of a $\Tvir>10^4$ K halo, we calculate its star formation rate as
\begin{equation}
\frac{\deriv M_*}{\dt} = \frac{c_*M_g}{\tdyn} ,
\label{eqn:psfr}
\end{equation}
where $M_g = M_b - M_*$ is the gas mass and $c_*$ is a star formation efficiency.  We keep track of the stellar fraction $f_*$ for each particle in order to differentiate the total baryonic mass into gas and stars.  Furthermore, we distinguish between the first generation, PopIII stars and the second generation, PopII stars.  It is postulated and this is supported by hydrodynamic simulations \citep{2006OsheaNorman, 2006YOHA} that the first generation of stars forming from pristine primordial gas are very massive $M\sim 100\Msun$ and the stellar IMF at high redshifts is top-heavy compared to a Salpeter IMF.

We have a simple prescription for the formation of PopIII stars motivated by results from hydrodynamic simulations.  Once a halo has accreted enough mass such that its virial temperature exceeds $10^4$ K, star formation is turned on and the IMF is determined by the metallicity of the halo.  An initially metal-poor halo with metallicity less than some critical metallicity $Z_\PopIII$ first undergoes PopIII star formation for a duration of $t_\PopIII$ years after which it switches to PopII star formation.  Hydrodynamic simulations have found that once the metallicity reaches a value $Z_\PopIII=10^{-3.5} \Zsolar$, metal line cooling becomes efficient enough that giant molecular clouds can undergo more fragmentation and form less massive stars.  The halo self-enrichment time $t_\PopIII=20$ Myr takes into account the lifetime ($\sim3$ Myr) of a massive star and the time it takes for supernova feedback to traverse and enrich the interstellar medium (ISM) with metals produced by the stars.  Furthermore, a fraction $f_{\rm Z,esc}=10\%$ of the metals are assumed to escape into the intergalactic medium (IGM) and we propagate them assuming a constant average velocity $v_{\rm Z,IGM}=10$ km/s.  By tracking the metals, we can suppress PopIII star formation using a local metallicity rather than relying on a global, volume-averaged metallicity for the IGM.

For PopII stars with a Salpeter IMF, the number of ionizing photons per baryon of star formation is 5200, 4100, and 270 for the three frequency ranges $\nu>$ 13.61, 24.59, and 54.42 eV, respectively.  For PopIII stars with a top-heavy IMF, the corresponding numbers are 70000, 55000, and 3500 \citep{2002Schaerer, 2003Schaerer}.  The radiation escape fraction $f_{\gamma,{\rm esc}}$ is defined to be the fraction of ionizing photons which escapes the ISM.

\subsection{Radiative transfer}

\begin{figure}[t]
\center
\includegraphics[width=3.4in]{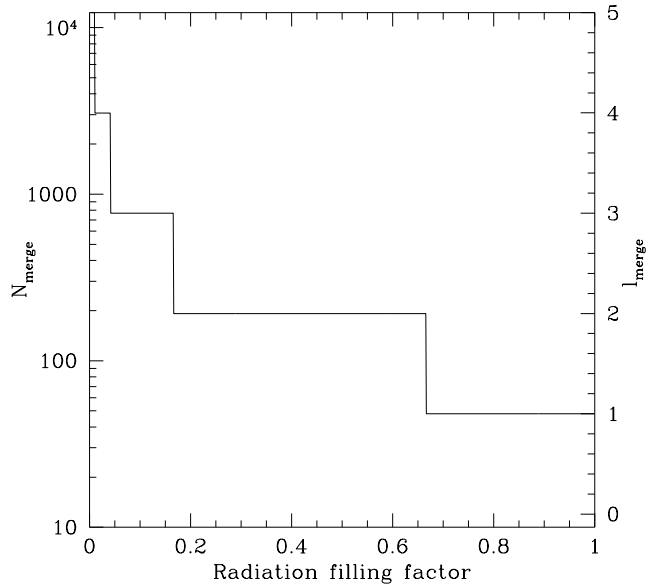}
\caption{In the adaptive ray merging scheme, $\Nmerge$ is the maximum number of rays with levels $l\geq \lmerge$ in a cell, but the total number of rays can be larger due to contributions from rays at lower levels,  which remain singular.  In the limit of complete reionization, the number of active rays will converge to a value of approximately $\Nmerge$ times the number of RT grid cells and the RT algorithm will scale as ${\cal O}(N)$.}
\label{fig:nmerge}
\end{figure}

Our unique approach to three-dimensional radiative transfer (RT) complements the high resolution and accuracy of the N-body simulations.  We have developed an adaptive ray tracing algorithm to accurately and efficiently propagate the complex radiation field generated by the many sources.  Furthermore, we calculate ionization and recombination for each individual particle, taking into account their geometrical cross-sections and volumes in order to correctly account for the clumping factors and self-shielding.  In all previous large volume simulations of reionization, the RT is calculated using low resolution density fields defined on a coarse grid, but small-scale information is lost with this approach.

First, we describe the set up for the RT calculations.  Particles, sources, and rays are collected on a grid with $\Nrt$ cells, where the mean number of particles per RT cell is $8^3$ and the number of PM cells per RT cell is $32^3$.  A source cell in the RT grid has a total star formation rate equal to the sum from all star-forming particles within the cell.  A radiation cell has a total number of photons equal to the sum from all rays intersecting it.  Each ray carries with it three numbers of photons for the frequency ranges (eV):  $13.61\leq\nu<24.59$, $24.59\leq\nu<54.42$, and $54.42\leq\nu<\infty$, respectively.  Particles are treated as individuals, but their differing photoionization rates will depend on the photon number density of the cell in which they belong to.  The HI, HeI, and HeII fractions for each particle are kept tracked of.  The radiative transfer time step $\ts_{\rm RT}$ is set by the light-crossing time for a RT cell and there are $\sim10-100$ RT time steps per PM time step.

We have implemented an adaptive ray tracing algorithm similar to that of \citet{2002AbelWandelt}, but have introduced a new merging scheme in order to avoid the ${\cal O}(N^2)$ scaling problem.  The ray tracing algorithm makes use of the HEALPix \citep{2002Healpix, 2005Healpix} equal-area pixelization scheme, which ensures that the rays uniformly cover the unit sphere.  We start with a base level $\lbase=1$ and cast $\Nbase=12\times4^{\lbase}=48$ rays per radiation source cell each radiative transfer time step.  In order to avoid artifacts arising from discreteness effects, one can randomly rotate the local coordinate system centered on each source cell, but we find it simpler and just as effective to randomize the origin of the local coordinate system within each source cell every time step.  Once a source cell becomes transparent, its 26 nearest neighbours will each, on average, have approximately 1.8 rays with radius $r=1$ RT grid unit.  We find that using a higher base level $\lbase=2$ and casting $\Nbase=192$ rays is unnecessary because of the randomization within source cells, the large number of RT times step per PM time step, and the large number of sources contributing to the radiation field.

As rays propagate, they adaptively split to ensure that a minimum number of rays, $N_{\rm min}$, from the same source intersect each radiation grid cell in the path each time step.  A ray with level $l$ and pixel number $\ipix$ splits into 4 daughter rays with levels $l+1$ and pixel numbers $4\ipix+n$ where $n$ runs from 0 to 3.  We use a value of $N_{\rm min}=1.8$ to be consistent with the base numbers for sources and this value is more than sufficient by the very same arguments given above.  A ray of radius $r$ in RT grid units will have a level,
\begin{equation}
l(r) = {\rm Integer}\left[\frac{\log(N_{\rm min}\pi r^2/3)}{\log(4)} \right] ,
\end{equation}
and since the Integer operator is always taken to round up, most cells will actually be intersected by more than $N_{min}$ rays from the same source each time step.  The adaptive splitting scheme provides excellent angular resolution for ray tracing from point sources, but it is computationally expensive because of the ${\cal O}(N^2)$ problem.

We have developed an adaptive ray merging scheme which converts the initially ${\cal O}(N^2)$ scaling to ${\cal O}(N)$ as the radiation filling factor approaches unity.  In our simulations, source cells can occupy up to $\sim10\%$ of the RT grid and therefore many rays will intersect any given cell at any given time.  Therefore, as the radiation filling factor increases, the angular resolution of rays entering and exiting cells can decrease without sacrificing much accuracy.  In our ray merging scheme, the number of rays in a cell at any given time is adaptively restricted with the simple formula:
\begin{equation}
\Nrayrh = \frac{\Nraymax}{\Nrt}\left(\frac{\Nrt}{\Nrad}\right) ,
\label{eqn:nrayrh}
\end{equation}
where $\Nraymax$ is the maximum number of rays and $\Nrad$ is the number of radiation cells intersected by at least one ray.  The ratio $\Nraymax/\Nrt$ sets the value we want in the limit of complete reionization and then it is divided by the radiation filling factor.  Eq.\ (\ref{eqn:nrayrh}) ensures that the number of active rays $\Nray$ will converge to $\Nraymax$ rather than growing quadratically with the size of the simulation.

The adaptive ray merging scheme is also implemented within the framework of HEALPix.  For each RT time step, we define the level for merging as,
\begin{equation}
\lmerge \equiv {\rm Integer}\left[\frac{\log(\Nrayrh/12)}{\log(4)}\right] ,
\end{equation}
where the Integer operator can round down, to the nearest, or up.  Rounding down is the conservative approach to satisfying Eq.\ (\ref{eqn:nrayrh}), but it unnecessarily lowers the angular resolution when the radiation filling factor is small.  Rounding up increases the angular resolution, but it may violate Eq.\ (\ref{eqn:nrayrh}) when the radiation filling factor is close to unity.  A good compromise, while still maintaining stability, is to round to the nearest integer.  For each cell, only rays with levels $l\geq \lmerge$ are considered for merging, while ones at lower levels remain singular.  The unit sphere is subdivided into $\Nmerge = 12\times4^{\lmerge}$ equal-area pixels and rays with $l\geq \lmerge$ are subdivided into $\Nmerge$ bins.  Rays with bin number $i_{\rm bin}$ are considered approximately parallel if they originate from rays of level $l=\lmerge$ which share the same pixel number $\ipix=i_{\rm bin}$.  Rays within a bin are merged into one and the total number of photons is conserved.  The new position, direction, and effective radius are calculated by taking averages weighted by the photon count of the individual rays.  

In Figure \ref{fig:nmerge}, the number $\Nmerge$ is plotted as a function of the radiation filling factor for $\Nrayrh=64$.  Note that $\Nmerge$ is the maximum number of rays with levels $l\geq \lmerge$ in a cell, but the total number of rays can be larger due to contributions from rays at lower levels.  For the majority of the simulation where the radiation filling factor is $<2/3$, $\Nmerge\geq 4\Nbase$ and the angular resolution in any given cell is better than the angular resolution of base rays in the source cells themselves.  We have experimented with higher angular resolution by increasing $\lbase$ to 2 and $\Nrayrh$ to 256, which comes with the extra cost of a factor of 4 in both computational work and memory, but find no significant effects on the radiation field.  

Rays are collected on the grid each RT time step using the nearest grid point (NGP) assignment scheme.  This mapping is robust because of the large number of rays intersecting any given cell and the large number of RT time steps per PM time step.  Our ray tracing algorithm shares several similarities to that of \citet{2007McQuinn}, who modified a previous algorithm \citep{2001Sokasian, 2003Sokasian}.  The algorithms have several commons, including how rays are cast, split, and mapped to the grid, but the main difference is how the problem of too many rays is solved.  They consider a simple approach where rays no longer split after traveling a certain distance.  The critical distance is chosen adaptively depending on the relative luminosity of the ray to the total luminosity of all rays in the box.  However, this isotropic value is not optimal for HII regions, which are quite anisotropic.  Our adaptive ray merging scheme is a more general approach and the ray tracing algorithm can be applied to many problems in astrophysics.

Once the photon density field is updated on the grid, we calculate the number of ionizations and recombinations for each individual particle rather than using the cell-averaged hydrogen and helium densities.  This approach maximizes the resolution and prevents the clumping factors from being severely underestimated.  Furthermore, working with the particles directly gives us the advantage of being able to account for self-shielding.  Star-forming halos which are sources will be ionized from the inside out, but the structure which are sinks will generally be ionized from the outside in.  To allow for self-shielding, the particles within each RT cell are sorted from lowest to highest density and the photons are absorbed by particles in this order.

The equations of ionization evolution for hydrogen and helium are:
\begin{align}
\frac{\deriv n_\HI}{\dt}	& = \alpha_\HII n_{\rm e} n_\HI - \Gamma_\HI n_\HI  ,\\
\frac{\deriv n_\HeI}{\dt}	& = \alpha_\HeII n_{\rm e} n_\HeII - \Gamma_\HeI n_\HeI ,\\
\frac{\deriv n_\HeIII}{\dt}	& = -\alpha_\HeIII n_{\rm e} n_\HeIII + \Gamma_\HeII n_\HeII ,
\end{align}
where $\alpha$ are the recombination rates and $\Gamma$ are the photoionization rates.  Consider a particle with volume $\Vp$, cross-section $\Ap$, and length $\lp$ in a RT cell where the number of photons per unit volume per unit frequency at frequency $\nu$ is $\eta_\nu$ (cm$^{-3}$ Hz$^{-1}$).  The rate of change of neutral hydrogen in the particle due to photoionization is given by
\begin{equation}
\dot{n}{_\gHI}\equiv \Gamma_\HI n_\HI = \int^\infty_{\nu_\HI} \frac{w_\fHI \dot{N}_\nu(1-e^{-\tau_\fHI})}{\Vp}\dnu ,
\label{eqn:photorate}
\end{equation}
where $\dot{N}_\nu = \eta_\nu \Ap c$ (s$^{-1}$ Hz$^{-1}$) is the number of photons per unit time per unit frequency passing through the particle, $\tau_\fHI = n_\HI \sigma_\fHI \lp$ is the optical depth for absorption by neutral hydrogen, and the weight,
\begin{equation}
w_\fHI = \frac{n_\HI\sigma_\fHI}{n_\HI\sigma_\fHI+n_\HeI\sigma_\fHeI+n_\HeII\sigma_\fHeII}
\end{equation}
takes into account the competition between HI, HeI, and HeII for the same photons.  Source particles are set to be fully ionized since the radiation escape fraction is defined to be that which escapes the ISM.  Eq.~(\ref{eqn:photorate}) is similar to Eq.~(15) from \citet{1999Abel}, except we have generalized it to account for a range in frequencies.  The rates of change for HeI and HeII by photoionization similarly follow that for HI.

Since the equations of ionization equilibrium are stiff, the time integration is computed using a stable implicit scheme rather than an explicit one.  Time integration is performed over the RT time step, but often subcycling in steps of the recombination time is required since the latter time scale is generally smaller.  The rays have their photon numbers reduced in proportion to the fraction of consumed photons throughout the cell.  In a cell where $<0.01\%$ of the photons remain unconsumed, we delete all the rays within in order to be computationally efficient and this has negligible effect on the results.

\section{Simulations}
\label{sec:simulations}

\begin{deluxetable*}{lrrrrrll}
\tabletypesize{\footnotesize}
\tablecaption{\label{tab:simparams} Simulation parameters}
\tablecolumns{8}
\tablehead{Model & $L$ (Mpc/$h$) & $N_p$ & $N_{PM}$ & $\Nrt$ & $\Nraymax$ & Stars & Comments} 
\startdata
L50a		& 50		& 1440$^3$	& 5760$^3$	& 180$^3$	& (1440$^3$)/8		& PopII 	& 
$f_{\gamma, {\rm esc}}=0.16$\\
L50b		& 50		& 1440$^3$	& 5760$^3$	& 180$^3$	& (1440$^3$)/8		& PopII \& PopIII	& 
$f_{\gamma, {\rm esc}}=0.15$, $t_\PopIII=20$ Myr, $v_{\rm Z,IGM}=10$ km/s\\
L100a	& 100	& 2880$^3$	& 11520$^3$	& 360$^3$	& N/A			& PopII \& PopIII	& 
Shin et al. (2007) with alternative RT algorithm\\
L100b	& 100	& 2880$^3$	& 11520$^3$	& 360$^3$	& (2880$^3$)/8		& PopII \& PopIII	& 
Trac et al. (2007)
\enddata
\end{deluxetable*}

\begin{figure*}
\center
\includegraphics{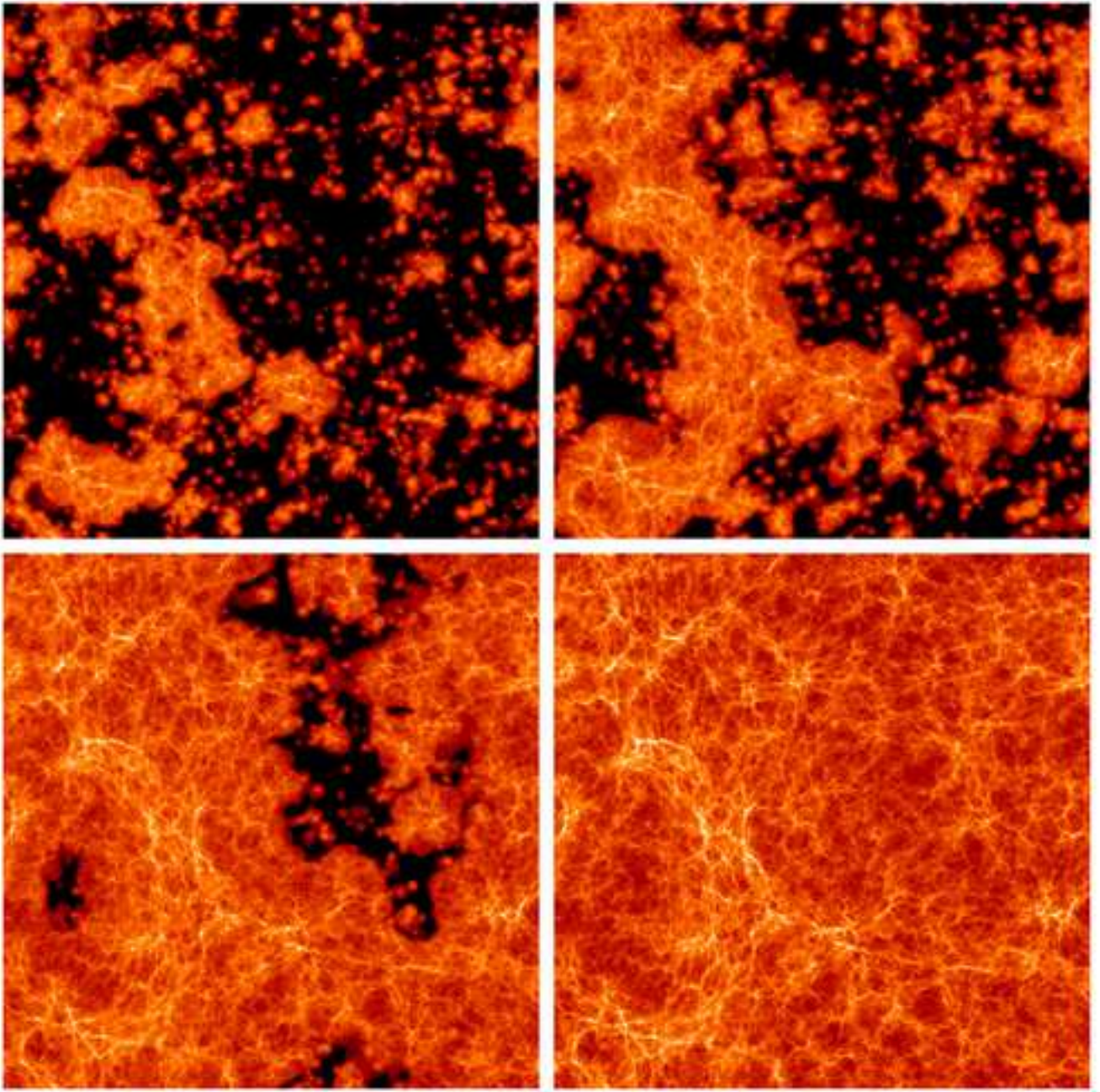}
\caption{Low resolution image showing the HII density in 1 Mpc/$h$ deep slices through the L50b simulation box at redshifts $z=$ 9, 8, 7, and 6.  The volume-weighted mean HI fraction are $f_\HI=$ 0.67, 0.46, 0.12, and $1.5\times10^{-4}$, respectively.  Higher resolution images can be found at http://www.astro.princeton.edu/$\sim$htrac/reionization.html.}
\label{fig:HIIevolution}
\end{figure*}

The hybrid N-body plus RT simulations of reionization were run with the following cosmological parameters:  $\Om=0.26$, $\Ol=0.74$, $\Ob=0.044$, $h=0.72$, $\sigma_8=0.77$, and $n_s=0.95$.  The N-body initial conditions were generated for an initial redshift $z=300$ where the matter power spectrum is still linear and the real space density field has $|\delta_{\max}|<1$ and $\sigma_\delta\sim0.03$.  The initial matter transfer function was computed with CMBFAST \citep{1996SeljakZaldarriagaCMBFAST} and the Zeldovich approximation was used to calculate the displacement and velocity for the particles.

Table \ref{tab:simparams} lists the parameters for the two test simulations, each with 3 billion particles in a 50 Mpc/$h$ box.  For the PM calculations, the comoving grid spacing is $\Delta x_{\rm PM} = 8.68$ kpc/$h$ and the particle mass resolution is $m_{\rm p} = 3.02\times10^6\Msun/h$, which is a substantial improvement over all previous large-scale simulations of reionization.  Halos are identified down to a minimum mass of $M_{\rm min} = 6\times10^7\Msun/h$.  The RT grid has a comoving grid spacing of $\Delta x_{\rm RT} = 278$ kpc/$h$ and sources and photons are collected at this resolution.  However, particles are treated individually for the ionization and recombination calculations, allowing us to probe the clumping factors down to scales of $\Delta x_{\rm PM}$ rather than $\Delta x_{\rm RT}$.  Figure \ref{fig:HIIevolution} is a sample visualization of the HII density evolution with PopIII stars.

The simulations were run at the National Center for Supercomputing Applications (NCSA) on a shared-memory SGI Altix with Itanium 2 processors.  Each test simulation used 128 processors and 360 GB of memory and required approximately 12000 cpu hours to complete.  Furthermore, we have completed a large simulation with 24 billion particles in a 100 Mpc/$h$ box.  The L100a simulation, run with an alternative RT algorithm, used 512 processors, 2 TB of memory, and approximately 80000 cpu hours.  Detailed results will be presented in a companion paper (Shin et al.~2007).

\section{Results}
\label{sec:results}

\subsection{Halo mass functions}
\label{sec:halos}

\begin{figure}
\center
\includegraphics[width=3.4in]{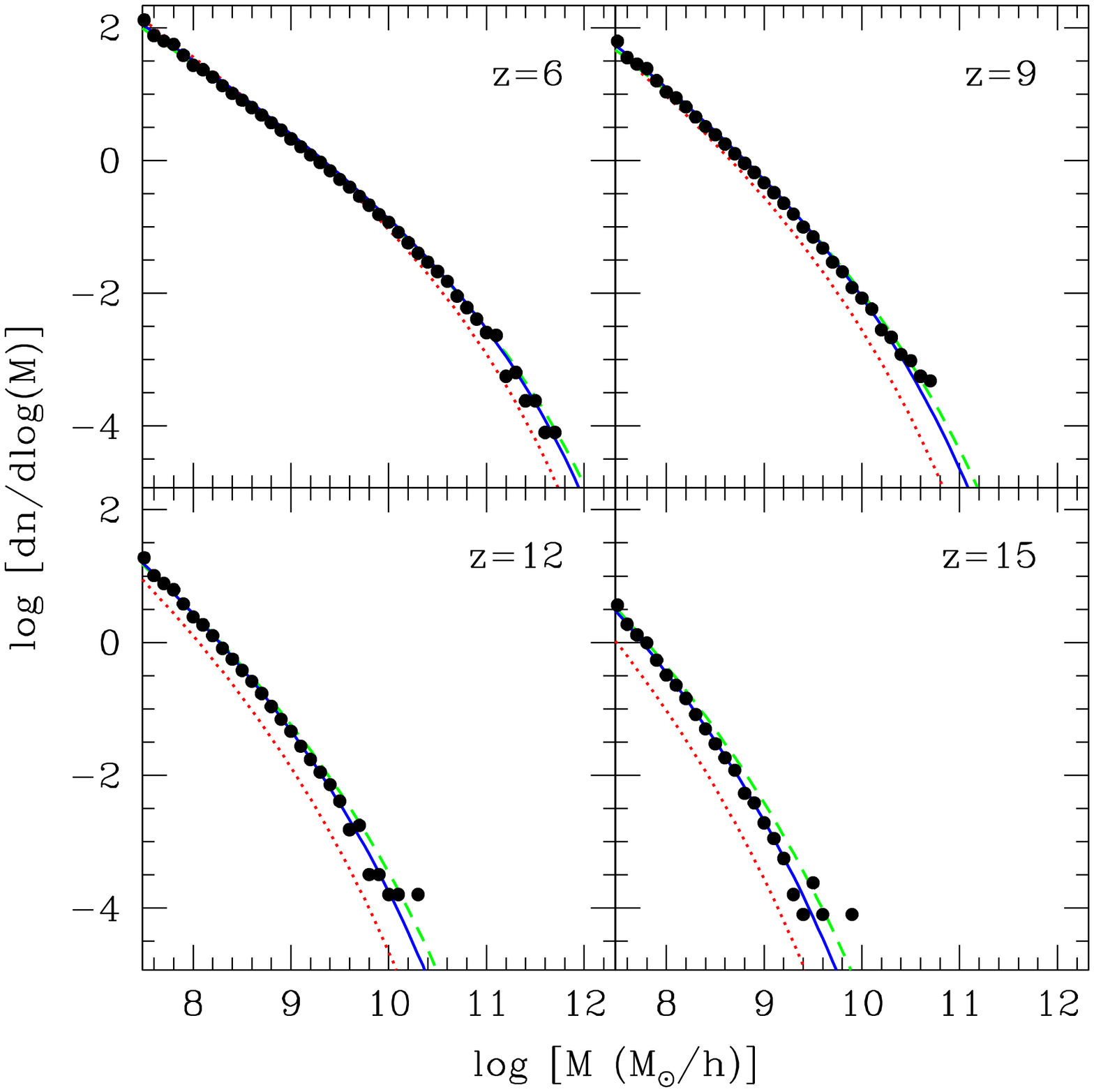}
\caption{Friends-of-friends differential mass functions of dark matter halos.  The halos are identified using a standard linking length of 0.2 times the mean interparticle spacing.  The simulation results are best fit by \citet[][blue, solid]{2006WarrenAHT} at all redshifts.  \citet[][green, dashed]{1999ShethTormen} tends to give higher abundances at higher redshifts, while \citet[][red, dotted]{1974PressSchechter} underpredicts the abundances at all redshifts.}
\label{fig:massfnc1}
\end{figure}

\begin{figure}
\center
\includegraphics[width=3.4in]{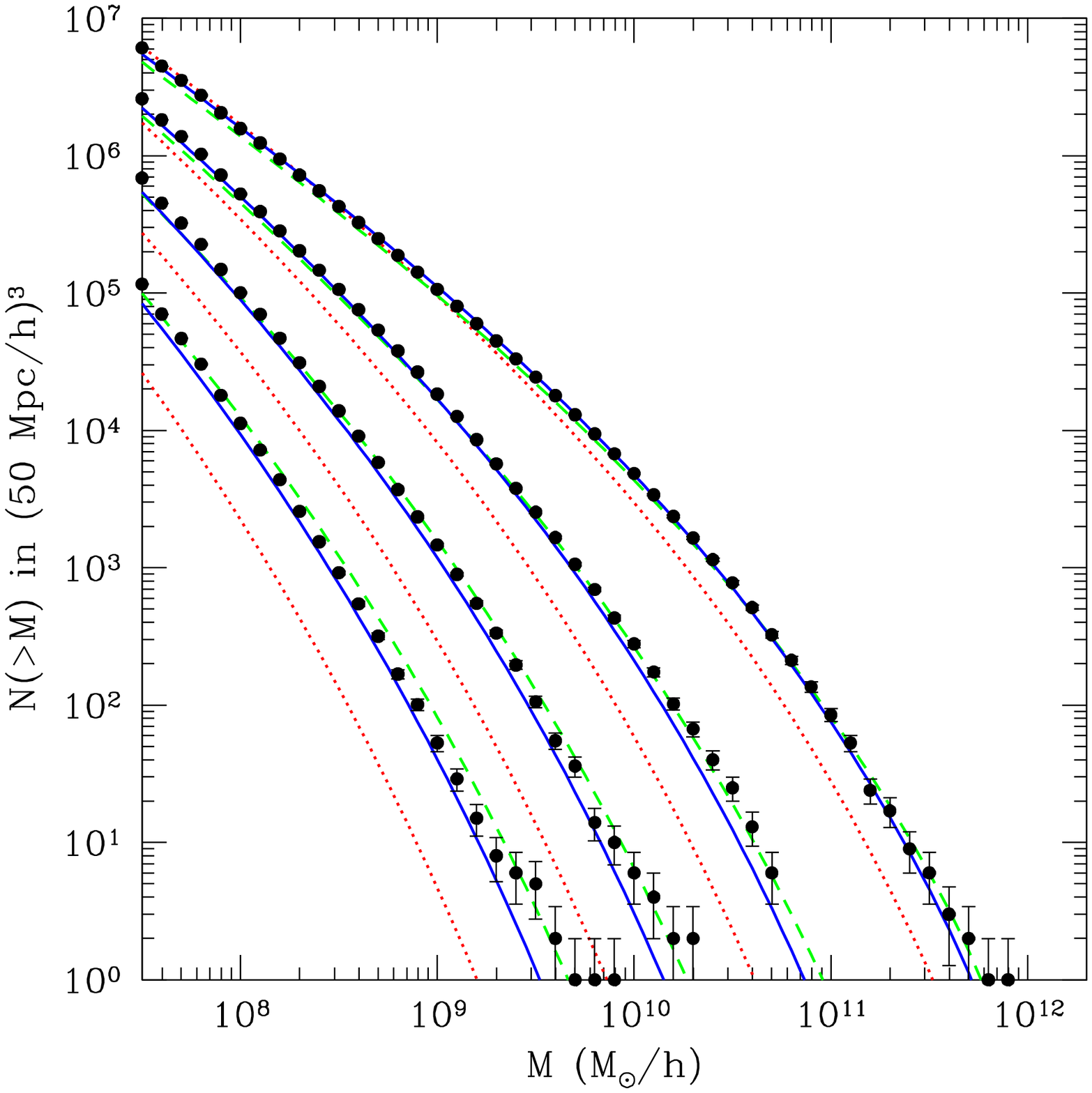}
\caption{Friends-of-friends cumulative mass functions of dark matter halos at $z$ = (6, 9, 12, 15) with Poisson error bars.  The \citet[][blue, solid]{2006WarrenAHT} prediction provides a better fit than both \citet[][red, dotted]{1974PressSchechter} and \citet[][green, dashed]{1999ShethTormen}.}
\label{fig:massfnc2}
\end{figure}

We identified dark matter halos using a friends-of-friends (FoF) algorithm with a standard linking length of $b=0.2$ times the mean interparticle spacing.  Figure \ref{fig:massfnc1} shows the differential mass function d$n$/d$\log(M)$, where $n(M,z)$ is the comoving number density of halos with mass less than $M$ at redshift $z$.  Figure \ref{fig:massfnc2} shows the cumulative mass function $N(M,z)$, defined as the number of halos with mass greater than $M$ at redshift $z$.  The FoF results from the L50 simulations are in best agreement with \citet{2006WarrenAHT}, who conducted a careful measurement of the mass function of dark matter halos with a suite of high-resolution N-body simulations.  The \citet[ST;][]{1999ShethTormen} model also provides a good fit, but still differs by several tens of percents, with larger differences at higher redshifts.  The original \citet[PS;][]{1974PressSchechter} model underpredicts the halo abundance at all our redshifts.  Our results are consistent with recent work on the mass function of high redshift dark matter halos \citep{2007ReedBFJT, 2007LukicHHBR, 2007CohnWhite} and are further confirmation that standard PS underpredicts the abundance of high redshift halos, particularly at the high mass end.

In earlier preliminary work, we used simulations of the same size but with a late starting redshift of $z=60$.  We also previously used a spherical overdensity (SO) algorithm with a characteristic density chosen to be 200 times the critical density, similar to that described in \citet{1994LaceyCole}.  The SO results appeared to be in better agreement with PS, particularly at higher redshifts and higher masses.  However, we were systematically under-resolving halos because our starting redshift was too late to accurately capture the nonlinear gravitational collapse.  \citet{2007ReedBFJT} have suggested that simulations must start $\sim10-20$ expansion factors before the redshift at which results converge.  Another systematic effect came from our SO definition of halo mass.  M200 is generally smaller than MFoF and a mass function constructed with the former will have a lower amplitude than the latter at a fixed mass.  \citet{2007CohnWhite} have quantified the difference in the mass function when two FoF linking lengths of 0.2 and 0.168 are used.  While the mass function for the larger linking length is in good agreement with ST, the results for the smaller linking length are better fit by PS but still with discernible differences.

\subsection{Halo clustering}

\begin{figure}
\center
\includegraphics[width=3.4in]{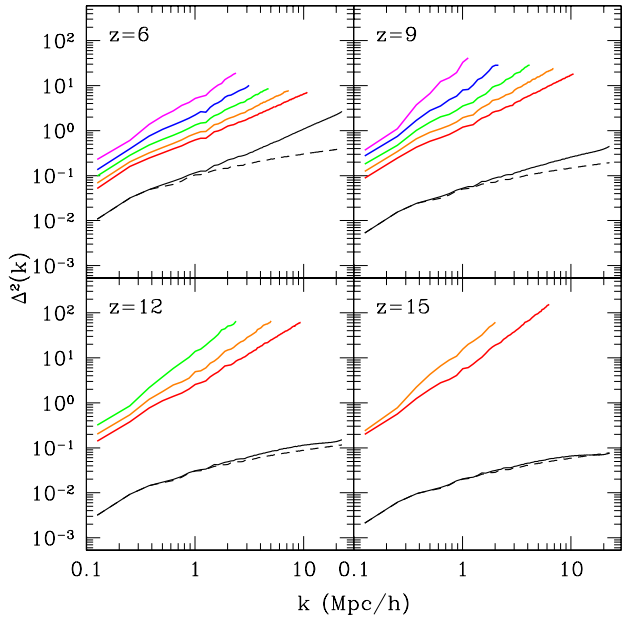}
\caption{Dark matter halo power spectra are shown for five logarithmic mass ($\Msun/h)$ bins (from bottom to top):  8.0-8.5 (red), 8.5-9.0 (orange), 9.0-9.5 (green), 9.5-10.0 (blue), and 10.0-11.0 (magenta).  The nonlinear (black, solid) and linearly extrapolated (black, dashed) matter power spectra are also shown for comparison.}
\label{fig:pkhalo}
\end{figure}

\begin{figure}
\center
\includegraphics[width=3.4in]{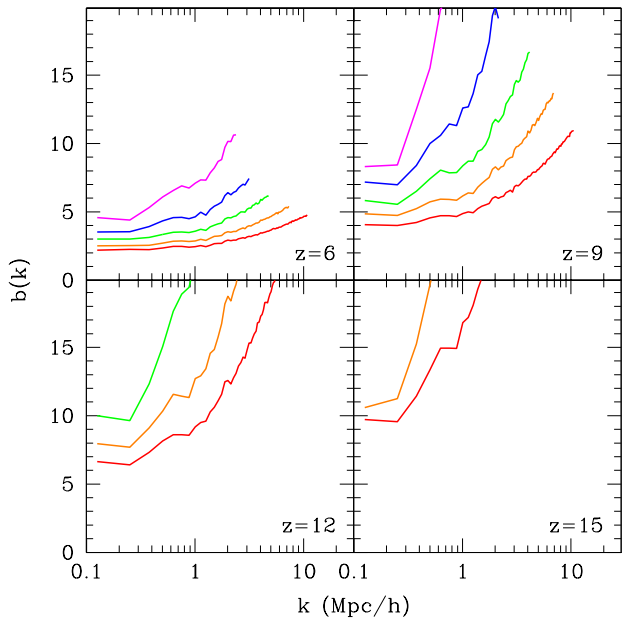}
\caption{Dark matter halo bias are shown for five logarithmic mass ($\Msun/h)$ bins (from bottom to top):  8.0-8.5 (red), 8.5-9.0 (orange), 9.0-9.5 (green), 9.5-10.0 (blue), and 10.0-11.0 (magenta).  The bias increases with mass and with redshift and is only linear on scales where $\delta_{\rm h}$ is small.}
\label{fig:bkhalo}
\end{figure}

\citet{2004BarkanaLoeb} have emphasized that the high redshift halos where photoionizing sources reside are highly biased and this is clearly seen in our simulations.  In Figure \ref{fig:pkhalo}, we plot the dimensionless halo power spectrum, 
\begin{equation}
\Delta_{\rm hh}^2(k,M,z)\equiv\frac{k^3}{2\pi^2} \langle |\delta_{\rm h}(k,M,z)|^2\rangle ,
\end{equation}
for wavenumbers $k$ where the signal to noise is greater than unity.  The Poisson noise is subtracted by removing the white noise power due to all self-pairs.  All halo spectra resemble power-laws, regardless of mass and redshift, and there appears to be an inverse relationship between the effective slope and the halo number density.  This suggests that the nonlinear clustering of dark matter halos can be described with a self-similar parametrization.  We will quantify this relationship in upcoming work.  

In Figure \ref{fig:bkhalo}, we plot the halo bias,
\begin{equation}
b(k,M,z) = \sqrt{\frac{\Delta_{\rm hh}^2(k,M,z)}{\Delta_{\rm lin}^2(k,z)}}
\end{equation}
where $\Delta_{\rm lin}^2(k,z)$ is the linear matter power spectrum extrapolated to redshift $z$.  The halo bias increases with mass and with redshift and is only linear on scales where $\delta_{\rm h}$ is small.  The linear bias has been derived for the PS model \citep{1996MoWhite} and for the ST model \citep{1999ShethTormen}.  PS predicts higher bias than the simulations because in this model, halos are more rare at a fixed mass and redshift.  On the contrary, the ST prediction gives slightly lower values than our results.

Since the linear bias is inversely related to the halo number density, the derivative ${\rm d}b/{\rm d}M$ increases rapidly with mass because of the exponential decline in the abundance of massive halos.  This rapid change is particularly more pronounced at higher redshifts.  The mass-dependent bias of halos has important implications for the clustering of sources and HII regions.  \citet{2006Iliev} were only able to resolve halos with $M>2.5\times10^9\Msun$, which is a factor of 30 times larger than our minimum halo mass.  In their simulation, reionization ended at $z\approx12$ and at this redshift, their power spectrum of all resolved halos is approximately a factor of 3 larger than ours.  The relative bias will be even larger at higher redshifts.  Recent work \citep{2007Zahn, 2007McQuinn} have pointed out that this leads to a very different picture of reionization, characterized by relatively fewer but larger and more spherical HII regions.  Therefore, simulations of reionization must resolve halos down to a minimum mass where the virial temperature is $\sim10^4$ K in order to correctly capture the clustering of sources and HII regions.

\subsection{Star formation}
\label{sec:starformation}

\begin{figure}
\center
\includegraphics[width=3.4in]{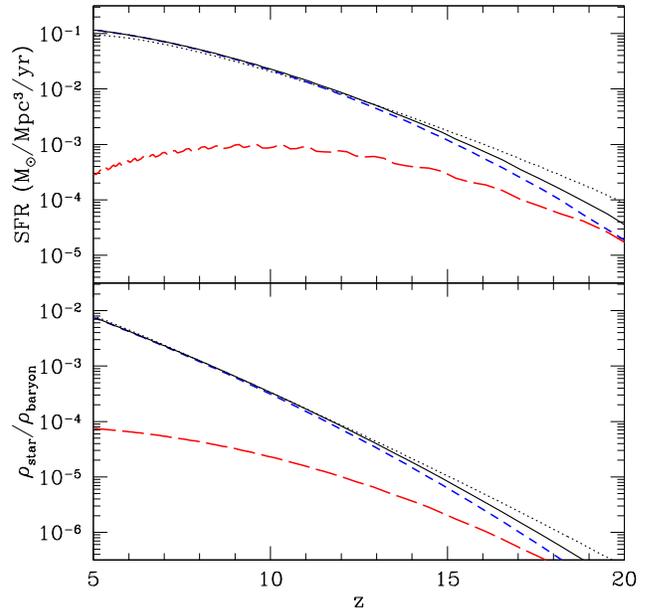}
\caption{The top panel shows the comoving star formation rate (SFR) from the L50b simulation with PopIII stars.  The total SFR (black, solid), from PopIII (red, long-dashed) and PopII (blue, short-dashed) stars, is in good agreement with the semi-analytical model of \citet[][black, dotted]{2003HernquistSpringel}.  The bottom panel shows the cumulative stellar density as a fraction of the mean baryon density.}
\label{fig:sfr}
\end{figure}

\begin{figure}
\center
\includegraphics[width=3.4in]{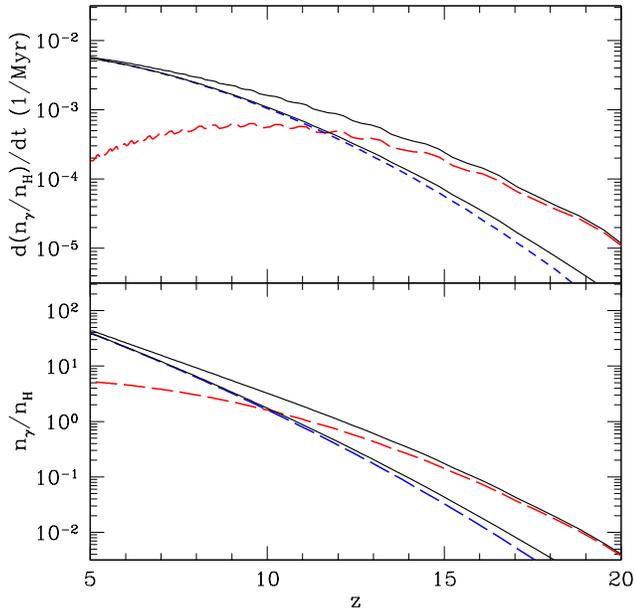}
\caption{The top panel shows the comoving photon production rate ${\rm d}(\ngam/\nH)/\dt$ of photons with energies $>13.61$ eV.  The total rates from the L50a and L50b simulations are given by the lower and upper (black, solid) curves, respectively.  In the L50b simulation, the PopIII (red, long-dashed) stars contribute a larger fraction to the total rate than the PopII (blue, short-dashed) stars for $z\gtrsim12$.  The bottom panel shows the cumulative photon density as a fraction of the mean hydrogen density.  In the L50b simulation, the photon budget is dominated by PopIII stars for $z\gtrsim10$.}
\label{fig:photon}
\end{figure}

In principle, the cosmic star formation rate (SFR) can be modelled using hydrodynamic simulations \citep[e.g.][]{2003SpringelHernquist, 2004NCHOS} and calibrated against low redshift observations.  \citet{2003HernquistSpringel} have derived an analytical formula for the SFR by considering the mass within the cooling radius of collapsed halos.  The free parameters of the model are fitted using results from hydrodynamic simulations.  In our simulations, the shape of the curve $\sfr$(z) agrees well with theirs and we choose the star formation efficiency $c_*=0.03$ to match their normalization, corrected for our cosmology.  It is encouraging that our star formation history is very similar to that obtained in hydrodynamic simulations and semi-analytic models.  Note that in our simulations, the number of escaped ionizing photons depends on the product $c_*f_{\gamma, {\rm esc}}$. 
While the star formation efficiency and the radiation escape fraction are degenerate in this regard, it is still important to model their values individually since other calculations will require different combinations of these two parameters.  For instance, in scatter calculations for high redshift Lyman alpha emitters \citep{2006Tasitsiomi}, the product $c_*(1-f_{\gamma, {\rm esc}})$ is also important.

Figure \ref{fig:sfr} shows the comoving SFR and the cumulative stellar density as a function of redshift.  The L50a and L50b simulations have the same total SFR, but the latter differentiates between PopII and PopIII stars.  While the stellar density of PopII stars is generally larger than that of PopIII stars, the number density of photoionizing photons produced by the former does not exceed that of the latter until $z=10$, as shown in Figure \ref{fig:photon}.  We assume that PopIII stars have a top-heavy IMF and produce a factor of 13.5 more photoionizing photons per unit stellar mass.  In the L50b simulation, the photon production rate ${\rm d}\ngam/\dt$ changes more gradually with redshift.  We show that this leads to a more extended reionization epoch in \S\ref{sec:ionfrac} and to a larger Thomson optical depth in \S\ref{sec:opticaldepth} when the redshift of complete reionization is fixed.

The SFR$_\PopIII$ reaches a broad maximum value of $\sim10^{-3}\Msun/{\rm Mpc^3/yr}$ at $z\approx10$, which is consistent with semi-analytical models \citep{2004YBH, 2006WyitheCen}.  We find that the decline of PopIII stars at lower redshifts is mainly due to the following reason.  In our simulations, PopIII stars form in metal-poor halos which have just accreted enough mass such that the virial temperature exceeds $10^4$ K.  However, the cooling mass $\Mvir(\Tvir=10^4\ {\rm K})$ is increasing as the redshift decreases (see Fig.~\ref{fig:msf}) and the formation rate of metal-poor halos with this mass is decreasing in time.  We also find that the amplitude of the SFR$_\PopIII$ is more strongly dependent on the halo self-enrichment time $t_\PopIII$ than on the metal enrichment of the IGM, which is controlled by the escape fraction of metals $f_{\rm Z,esc}$ and the average propagation speed of metals $v_{\rm Z,IGM}$.  The relative contribution of PopIII stars to the total SFR is very uncertain and the parameters of the model will be studied in upcoming work.

\citet{2007McQuinn} considered several models for the source efficiency, defined to be proportional to $\dot{M}_*/M$ for star forming halos of mass $M$.  They considered cases where the source efficiency is independent of $M$ or scales as $M^{-2/3}$ or $M^{2/3}$, but with no redshift dependence.  In our star formation prescription, the local star formation rate is proportional to the gas density and inversely proportional to the dynamical time (see Eq.~\ref{eqn:psfr}).  When the redshift is kept fixed, the source efficiency has negligible dependence on mass because the halos have very similar density profiles.  They are defined to have the same average density of 200 times the critical density and have similar concentration parameters (see Eq.~\ref{eqn:concentration}).  However, for a fixed mass, the source efficiency scales as $(1+z)^{3/2}$ at high redshifts because of the inverse dependence on the dynamical time.  

The source efficiency can be a complicated function of redshift, mass, and environment.  Feedback from photoionization and supernovae can alter the SFR locally, particularly in the lower mass halos.  \citet{2006IMSP} have used a toy model to study the suppression of star formation during the reionization epoch.  They argue that photoheating will raise the Jeans mass scale for gravitational collapse and thereby reduce or halt star formation in lower mass halos.  Feedback effects are actually much more complicated.  Photoionization produces electrons, which are catalysts in H2 formation, and this can lead to enhanced cooling.  Supernovae will inject energy into the ISM and IGM, but the metal enrichment can lead to additional cooling.  These nonlinear effects are best investigated with high resolution, small volume hydrodynamic simulations \citep[e.g.][]{2000Gnedin}.  They can then be incorporated into the star formation prescription for large volume simulations of reionization.

\subsection{Ionization fractions}
\label{sec:ionfrac}

\begin{figure*}
\center
\includegraphics[width=7in]{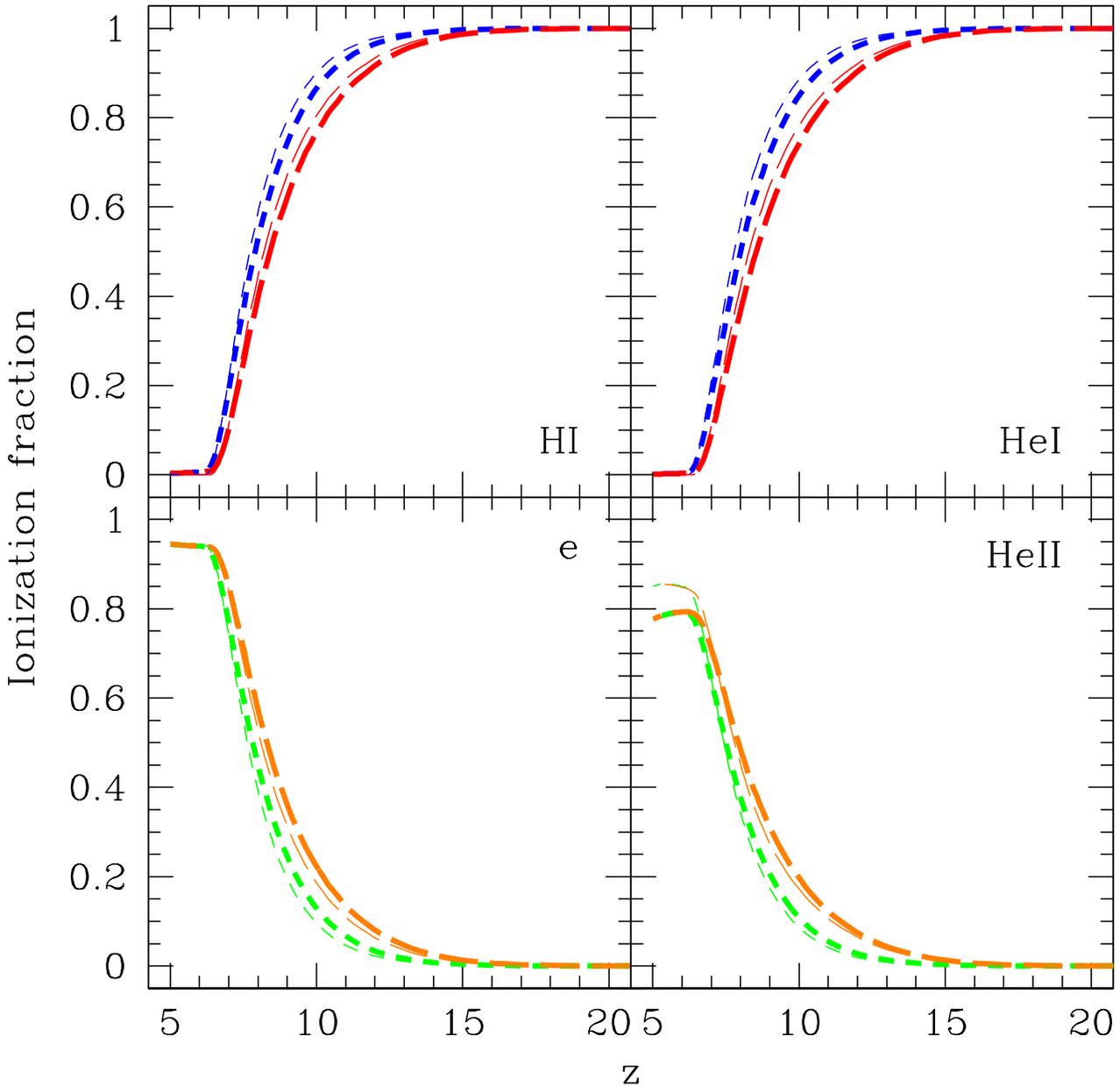}
\caption{Mass (thicker lines) and volume (thinner lines) weighted ionization fractions.  In the top panels, we show the HI and HeI fractions from simulations with (red, long-dashed) and without (blue, short-dashed) PopIII stars.  HI and HeI are ionized similarly, both spatially and with redshift, because they have similar absorption and recombination coefficients.  In the bottom panels, we show the electron and HeII fractions with (orange, long-dashed) and without (green, short-dashed) PopIII stars.  Reionization commences earlier and the duration is longer with the addition of PopIII stars in the L50b simulation.}
\label{fig:ionization}
\end{figure*}

\begin{figure}
\center
\includegraphics[width=3.4in]{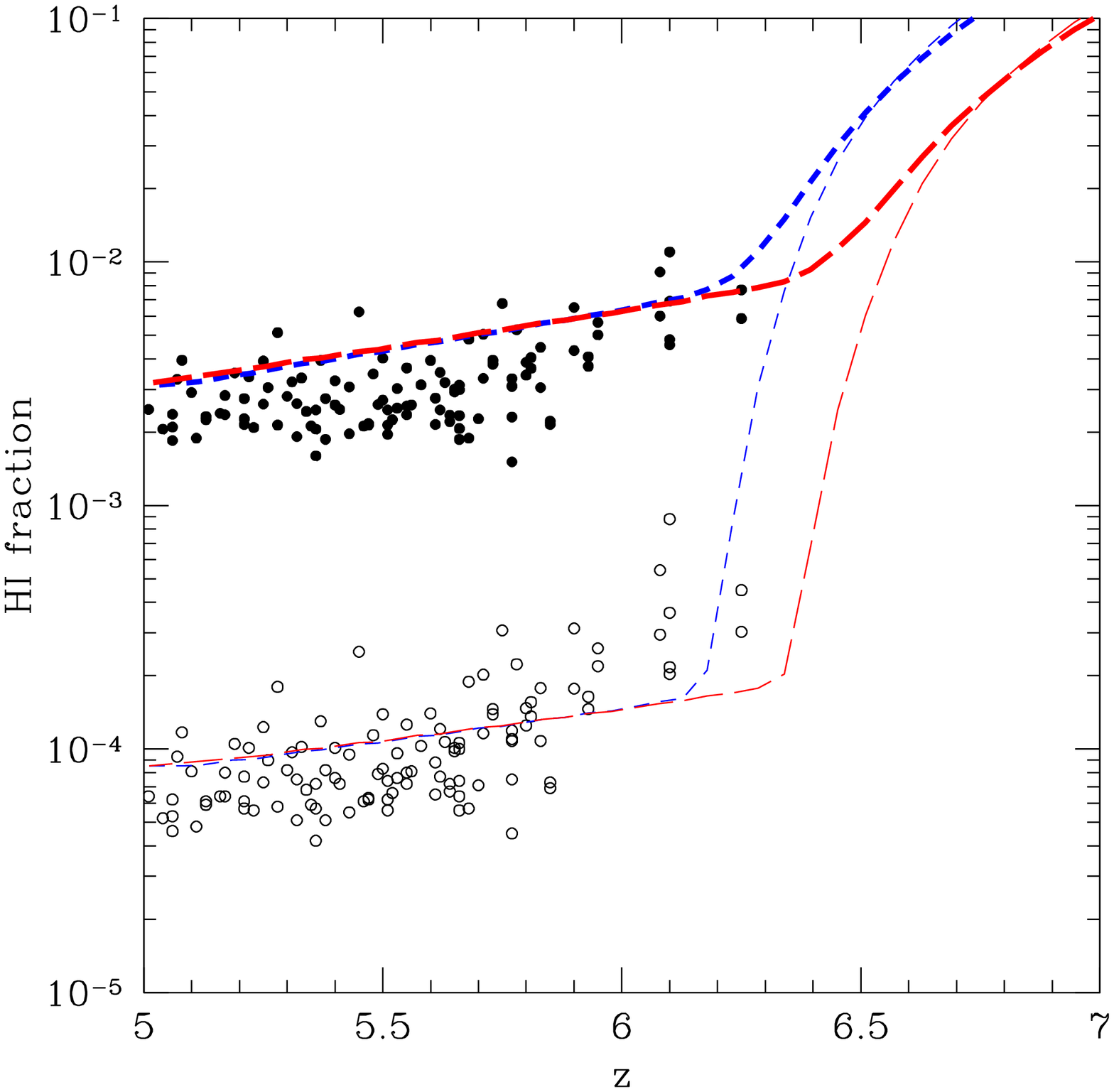}
\caption{Mass (upper, thicker lines) and volume (lower, thinner lines) weighted HI fractions with (red, long-dashed) and without (blue, short-dashed) PopIII stars.  The simulated residual HI fractions are in good agreement with measurements from 19 high redshift quasars in the SDSS \citep{2006Fan}.}
\label{fig:residualHI}
\end{figure}

We plot the mass and volume weighted ionization fractions for HI, HeI, HeII, and electrons in Figure \ref{fig:ionization}.  In order to have reionization end at $z\approx6.5$, we chose the L50a and L50b simulations to have radiation escape fractions of 16\% and 15\%, respectively.  Despite the lower escape fraction in the latter simulation, there are more ionizations at higher redshifts due to the presence of PopIII stars.  HI and HeI are ionized similarly, both spatially and with redshift, because they have similar absorption cross-sections and recombination coefficients.  Only a small fraction $\sim10\%-20\%$ of HeII is ionized by $z=5$, with more in the L50b simulation because the PopIII stars are assumed to have a top-heavy IMF with an effective spectrum which peaks at higher frequencies compared to PopII stars with a Salpeter IMF.

In Figure \ref{fig:residualHI}, we show that the computed mass and volume weighted residual HI fractions at $z\lesssim6.5$ are both in good agreement with the measurements from high redshift quasars in the SDSS \citep{2006Fan}.  \citet{2006GnedinFan} have shown that high resolution, small volume hydrodynamic simulations which resolve small-scale objects like Lyman limit systems, can correctly count the number of absorptions.  Figure 5 in their paper shows the agreement in the volume weighted HI fraction, but it is not obvious that the mass weighted fraction is also in agreement.  Their simulations have a maximum box length of 8 Mpc/$h$ and will give highly biased results for both sources and HII regions.  To date, no other numerical work involving large volume simulations have reported this agreement with observations.  In upcoming work, we will compare the HI optical depths \citep{1965GunnPeterson} since it is a more direct observational measurement than the neutral fraction.  The mean HI fractions calculated by \citet{2006Fan} are based on a model of the density distribution of the IGM by \citet{2000MiraldaEscude} and this model needs to be updated before a precise comparison can be performed.  Furthermore, we plan to compare the dark gap distributions \citep{2005PaschosNorman} with high redshift observations.  We can probe the response of the gas to reionization by modelling and varying the photoionization feedback in our baryonic prescription.

\subsection{Clumping factors}
\label{sec:clumpingfactors}

\begin{figure}
\center
\includegraphics[width=3.4in]{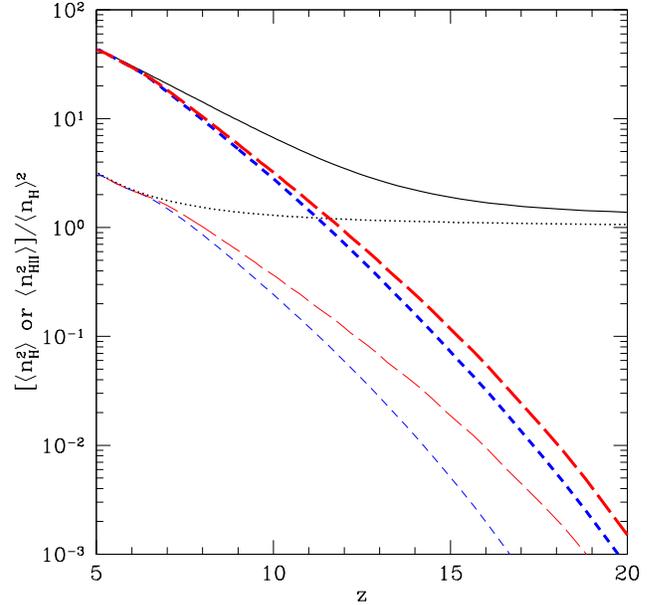}
\caption{Clumping factors for H and HII measured at high resolution with the particles (thicker) and at low resolution with the RT grid (thinner).  The $\nH$ clumping factors (black) are measured using all baryons.  The $\nHII$ clumping factors are measured using ionized gas from simulations with (red, long-dashed) and without (blue, short-dashed) PopIII stars.  All recombination calculations are performed at high resolution using the particles to prevent the clumping factors from being underestimated.} 
\label{fig:clumpingfactor}
\end{figure}

In Figure \ref{fig:clumpingfactor}, we compare the clumping factors measured at high resolution with the particles and at low resolution with the RT grid.  The particles can probe scales near the PM spacing $\Delta x_{\rm PM} = 8.68$ kpc/$h$, but the grid is smoothed on scales of $\Delta x_{\rm RT} = 278$ kpc/$h$.  We define the cosmic clumping factors for H and HII as,
\begin{align}
C_{\rm H} & \equiv \frac{\langle \nH^2\rangle}{\langle\nH\rangle^2} ,\\
C_\HII	& \equiv \frac{\langle\nHII^2\rangle}{\langle\nH\rangle^2} ,
\end{align}
where we normalize using the mean cosmic hydrogen density $\langle\nH\rangle$.  The clumping factors are calculated using the baryonic prescription which accounts for Jeans smoothing in identified collapsed halos.  Gas clumping factors derived using high resolution N-body simulations where the baryons are assumed to trace the dark matter at all scales are overestimates.

At high redshifts $z\gtrsim20$ when the density field is still highly linear, the H clumping factors from the particles and grid are close to unity.  As the redshift decreases, the abundance of collapsed halos increases and the universe becomes clumpier.  The values calculated from the particles reflect this change, but the grid values are underestimates.  At all redshifts prior to complete reionization, the HII clumping factors are more severely underestimated by the grid than for hydrogen.  Since HII regions originate from highly biased sources with large density contrast, the ionized gas will be very clumpy on scales much smaller than the grid smoothing scale.  By redshift $z=6$ when reionization is complete, the cosmic clumping factor is approximately 30, but the grid value is lower by a factor of 15.

Low resolution, large volume simulations have attempted to account for the subgrid clumping factor to improve the recombination calculations \citep[e.g.][]{2005KGH, 2006IMSP, 2007McQuinn}.  For a given density on the RT grid, they have applied an averaged correction, but have yet to account for the subgrid scatter.  We naturally account for scatter by directly working with the particles.

\subsection{Optical depth}
\label{sec:opticaldepth}

\begin{figure}
\center
\includegraphics[width=3.4in]{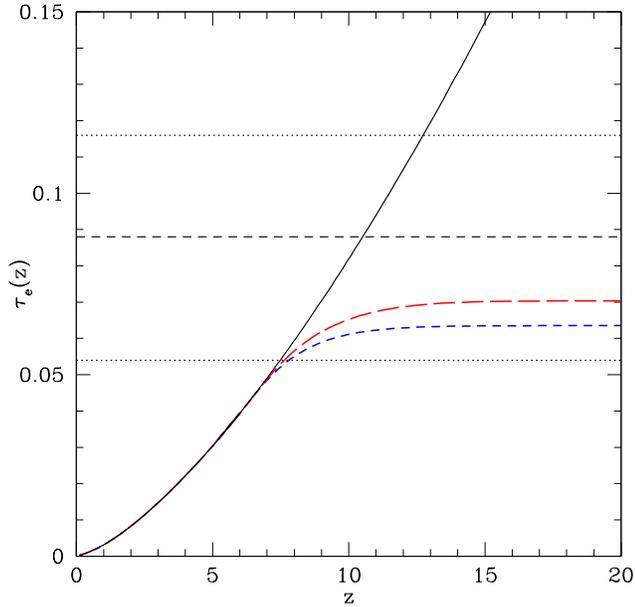}
\caption{The Thomson optical depth from simulations with (red, long-dashed) and without (blue, short-dashed) PopIII stars.  The three horizontal lines are the best-fit (dashed) and $1-\sigma$ (dotted) errors from third-year WMAP \citep[$\taue=0.088{+0.028}{-0.034}$;][]{2006PageWMAP3, 2007SpergelWMAP3}.  The (black, solid) curve $\taue(z)$ gives the optical depth for instantaneous reionization at redshift $z$.}
\label{fig:tau}
\end{figure}

The Thomson optical depth to electron scattering up to the redshift of recombination $z_{\rm rec}$ is given as,
\begin{equation}
\taue = \sigma_T\int^0_{z_{\rm rec}} \nelec(z) \frac{c\dt}{\dz}{\dz}
\end{equation}
where $\nelec(z)$ is the proper mean electron number density at redshift $z$.  WMAP polarization measurements with three years of data have yielded an optical depth $\taue=0.088^{+0.028}_{-0.034}$ \citep{2006PageWMAP3, 2007SpergelWMAP3}, which is a factor of 2 smaller than the first year result \citep{2003KogutWMAP}.  The best-fit value suggests that the universe was highly ionized by $z\sim10$, but the uncertainties are still large.  

In Figure \ref{fig:tau}, we show that the values calculated from the L50a and L50b simulations are consistent within $1-\sigma$, albeit on the low side.  In the latter simulation, $6\times10^{-5}$ of the baryons are turned into PopIII stars by redshift $z=6$.  The addition of this amount of PopIII stars has resulted in an increase of $\Delta\taue\approx0.007$ in the optical depth.  To match the best-fit value would require several times more PopIII stars at a fixed total SFR.  \citet{2006WyitheCen} have argued that the era of PopIII star formation can be significantly prolonged if mini-halos remain metal-poor until they merge into larger halos with virial temperatures above $10^4$ K where star formation is assumed to be more efficient.  If the redshift of complete HII overlap is kept fixed, then a more extended reionization epoch can be achieved by introducing more PopIII stars at higher redshifts.

Models of reionization with late overlap at redshift $z\sim6$ generally give lower values for the optical depth.  \citet{2007Zahn} recently obtained $\taue=0.06$ from a simulation where reionization ended at $z\approx6$.  However, their low resolution simulation will systematically underestimate the SFR at higher redshifts, resulting in a shorter reionization epoch and lower optical depth.  Simulations with earlier overlap as considered by \citet{2006Iliev, 2006IMSP} have given larger values which are closer to the current best-fit value.  However, it has not been shown whether simulations with early reionization can match observations of quasars at $5\lesssim z\lesssim 6.5$.  Previously in \S\ref{sec:ionfrac}, we discussed that the mass and volume weighted residual HI fractions in our simulations are both in good agreement with recent observations by \citet{2006Fan}.

We can account for all halos above the cooling mass $\Mvir(\Tvir=10^4\ {\rm K})$ for $z\lesssim15$, but miss relatively more of these halos as the cooling mass decreases with increasing redshift (see Fig.~\ref{fig:msf}).  Consequently, the computed optical depths may be underestimated, especially in the L50b simulation where PopIII stars make a comparatively larger contribution than PopII stars at higher redshifts.  Furthermore, a box size of 50 Mpc/$h$ is not quite large enough to calculate the optical depth accurately.  We have found that small boxes systematically underpredict the optical depth.  The artificial sudden overlap of the largest HII region with itself due to periodic boundary conditions \citep{2004BarkanaLoeb} results in a less extended reionization epoch and hence a lower value for the optical depth.  The two test simulations  are very useful for comparing the differences between reionization histories with and without PopIII stars, but larger simulations are required for detailed quantitative comparisons with observations.  In ongoing work using a large simulation with 24 billion particles in a 100 Mpc/$h$ box, we find that a modest increase in the SFR at high redshifts and in the total PopIII stellar density can give an optical depth of $\taue\approx0.09$ (Shin et al.~2007).

\subsection{Summary and Conclusions}

We have developed a new hybrid code to run large volume, high resolution radiative transfer simulations of cosmic reionization.  Two test simulations each with 3 billion particles and 400 million rays in 50 Mpc/$h$ boxes have been run to date to study reionization histories with and without PopIII stars.  Large simulations with 24 billion particles 100 Mpc/$h$ boxes will be presented in companion papers (Shin et al.~2007, Trac et al.~2007).

The RT calculations utilize an efficient adaptive ray tracing algorithm.  We use a ray splitting scheme \citep{2002AbelWandelt} to obtain robust angular resolution for ray tracing from point sources, but introduce a new ray merging scheme in order to handle the millions of sources.  Brute force ray tracing inherently scales as ${\cal O}(N^2)$, but with adaptive merging of co-parallel rays,  we converge on ${\cal O}(N)$ scaling as the radiation filling factor approaches unity.  This ray tracing algorithm can be combined with a cosmological hydrodynamic code to study the effects of reionization and a nonuniform radiation field, due to shadowing and shielding, on the high redshift Lyman alpha forest.

A main feature of our RT algorithm, which makes it unique compared to others used in large volume reionization simulations, is that the ionization and recombination calculations are performed on the particles rather than on a coarse grid.  We gain three major advantages by working with the particles directly.  First, clumping factors are calculated at high resolution down to scales of several comoving kpc rather than several hundred kpc.  Second, we account for the time-dependent photoionization cross-sections of neutral gas.  Third, we account for the self-shielding of radiation sinks.  

We use a particle-multi-mesh N-body code to track the gravitational evolution of the collisionless dark matter.  Collapsed dark matter halos are identified using a friends-of-friends algorithm.  The high redshift halos are strongly biased and linear theory breaks down at smaller $k$ than it does for the dark matter density field.  The halo spectra resemble power-laws, regardless of mass and redshift.  We will quantify the halo clustering in more detail with the larger simulations.

The two test simulations can identify halos down to a minimum mass of $6\times10^7\Msun/h$.  For $z<15$, we can account for all halos with virial temperatures $>10^4$ K and some mini-halos, but at higher redshifts, we miss some sources and all of the mini-halos.  As a result, we may be underestimating the reionization process and the Thomson optical depth.  Unresolved halos can be added using the method outlined in \citet{2007McQuinn}, but care must be taken to ensure that input halos at one redshift match simulated halos at a lower redshift.  They have applied the technique to post-processed data, but it is more difficult for us since the RT calculations are performed concurrently with the N-body calculations.

Baryons are assumed to trace the dark matter in low density regions, but a bias is introduced within collapsed halos to account for Jeans smoothing from gas pressure.  In the present version of the code, the gas is assumed to be in hydrostatic equilibrium and follows a polytropic equation of state.  The derived gas density and temperature profiles are then used to self-consistently calculate the gas clumping factors and star formation rates.  We have yet to take into account rotation, cooling, and feedback from photoionization and supernovae, but these effects will be incorporated into the baryonic prescription in upcoming work.

The star formation prescription closely resembles those used in hydrodynamic simulations.  Star formation is restricted to occur only in halos with virial temperatures above $10^4$ K and the local SFR is taken to be proportional to the gas density and inversely proportional to the dynamical time.  Our cosmic SFR is in good agreement with hydrodynamic simulations.  We find that the source efficiency is independent of mass for a fixed redshift, but is redshift dependent when the mass is fixed.  Photoionization, supernovae, and metal enrichment can alter the source efficiency and cosmic SFR.  These nonlinear effects should be investigated with high resolution, small volume hydrodynamic simulations and then incorporated into the star formation prescription for large volume simulations of reionization.

We have simulated two reionization models and have fixed the redshift of complete HII overlap to occur between $6\lesssim z\lesssim 6.5$.  Our results are consistent with the latest observations from SDSS and WMAP, both of which still have considerable uncertainties.  We emphasize that our values for the optical depth are for the two chosen reionization histories.  The actual star formation efficiency, number of ionizing photons per baryon, radiation escape fraction, and the clumpiness of the IGM at high redshifts are still unknown up to an order of magnitude.  Therefore, it is important to vary the parameter space in studies of reionization.  This can be efficiently accomplished by combining numerical and semi-analytical methods.

We will learn more about the epoch of reionization from additional observations of high redshift quasars in the SDSS and improved measurements of the Thomson optical depth from WMAP.  The next generation of observations, primarily high redshift surveys of Lyman alpha emitting galaxies, CMB measurements of KSZ effect from free electrons, and radio observations of the 21 cm radiation from neutral hydrogen, can provide significantly stronger constraints, but this demands much more accurate theoretical calculations.  Our approach for large volume, high resolution simulations can provide the accuracy and feasibility required for the task.

\acknowledgments

We thank Adam Lidz, Matt McQuinn, Min-Su Shin, Oliver Zahn, and Salman Habib for informative discussions.  We thank Lars Hernquist for the code to compute the semi-analytical SFR and Xiaohui Fan for the residual HI measurements from quasar absorptions.  This research is supported in part by grants AST-0407176 and NNG06GI09G.  HT is additionally supported in part by NASA grant LTSA-03-000-0090.  The simulations and analysis were performed at high performance computing facilities at the National Center for Supercomputing Applications (NCSA) and the Princeton Institute for Computational Science and Engineering (PICSciE).  We thank Dave McWilliams at NCSA and Bill Wichser at PICSciE for invaluable help with computing.

\bibliographystyle{apj}
\bibliography{astro}

\begin{thebibliography}{68}
\expandafter\ifx\csname natexlab\endcsname\relax\def\natexlab#1{#1}\fi

\bibitem[{{Abel} {et~al.}(1999){Abel}, {Norman}, \& {Madau}}]{1999Abel}
{Abel}, T., {Norman}, M.~L., \& {Madau}, P. 1999, \apj, 523, 66

\bibitem[{{Abel} \& {Wandelt}(2002)}]{2002AbelWandelt}
{Abel}, T., \& {Wandelt}, B.~D. 2002, \mnras, 330, L53

\bibitem[{{Barkana} \& {Loeb}(2001)}]{2001BarkanaLoeb}
{Barkana}, R., \& {Loeb}, A. 2001, \physrep, 349, 125

\bibitem[{{Barkana} \& {Loeb}(2004)}]{2004BarkanaLoeb}
---. 2004, \apj, 609, 474

\bibitem[{{Becker} {et~al.}(2001){Becker}, {Fan}, {White}, {Strauss},
  {Narayanan}, {Lupton}, {Gunn}, \& {Annis}}]{2001Becker}
{Becker}, R.~H., {Fan}, X., {White}, R.~L., {Strauss}, M.~A., {Narayanan},
  V.~K., {Lupton}, R.~H., {Gunn}, J.~E., \& {Annis}, J. 2001, \aj, 122, 2850

\bibitem[{{Bryan} \& {Norman}(1998)}]{1998BryanNorman}
{Bryan}, G.~L., \& {Norman}, M.~L. 1998, \apj, 495, 80

\bibitem[{{Cen}(1992)}]{1992Cen}
{Cen}, R. 1992, \apjs, 78, 341

\bibitem[{{Cen}(2003{\natexlab{a}})}]{2003Cena}
---. 2003{\natexlab{a}}, \apjl, 591, L5

\bibitem[{{Cen}(2003{\natexlab{b}})}]{2003Cenb}
---. 2003{\natexlab{b}}, \apj, 591, 12

\bibitem[{{Cen} \& {Ostriker}(1992)}]{1992CenOstriker}
{Cen}, R., \& {Ostriker}, J.~P. 1992, \apjl, 399, L113

\bibitem[{{Ciardi} {et~al.}(2003){Ciardi}, {Ferrara}, \& {White}}]{2003Ciardi}
{Ciardi}, B., {Ferrara}, A., \& {White}, S.~D.~M. 2003, \mnras, 344, L7

\bibitem[{{Cohn} \& {White}(2007)}]{2007CohnWhite}
{Cohn}, J.~D., \& {White}, M. 2007, ArXiv e-prints, 706

\bibitem[{{Dolag} {et~al.}(2004){Dolag}, {Bartelmann}, {Perrotta},
  {Baccigalupi}, {Moscardini}, {Meneghetti}, \& {Tormen}}]{2004Dolag}
{Dolag}, K., {Bartelmann}, M., {Perrotta}, F., {Baccigalupi}, C., {Moscardini},
  L., {Meneghetti}, M., \& {Tormen}, G. 2004, \aap, 416, 853

\bibitem[{{Fan} {et~al.}(2006{\natexlab{a}}){Fan}, {Carilli}, \&
  {Keating}}]{2006FCK}
{Fan}, X., {Carilli}, C.~L., \& {Keating}, B. 2006{\natexlab{a}}, \araa, 44,
  415

\bibitem[{{Fan} {et~al.}(2001){Fan}, {Narayanan}, {Lupton}, {Strauss}, {Knapp},
  {Becker}, {White}, {Pentericci}, {Leggett}, {Haiman}, {Gunn}, {Ivezi{\'c}},
  {Schneider}, {Anderson}, {Brinkmann}, {Bahcall}, {Connolly}, {Csabai}, {Doi},
  {Fukugita}, {Geballe}, {Grebel}, {Harbeck}, {Hennessy}, {Lamb}, {Miknaitis},
  {Munn}, {Nichol}, {Okamura}, {Pier}, {Prada}, {Richards}, {Szalay}, \&
  {York}}]{2001Fan}
{Fan}, X., {Narayanan}, V.~K., {Lupton}, R.~H., {Strauss}, M.~A., {Knapp},
  G.~R., {Becker}, R.~H., {White}, R.~L., {Pentericci}, L., {Leggett}, S.~K.,
  {Haiman}, Z., {Gunn}, J.~E., {Ivezi{\'c}}, {\v Z}., {Schneider}, D.~P.,
  {Anderson}, S.~F., {Brinkmann}, J., {Bahcall}, N.~A., {Connolly}, A.~J.,
  {Csabai}, I., {Doi}, M., {Fukugita}, M., {Geballe}, T., {Grebel}, E.~K.,
  {Harbeck}, D., {Hennessy}, G., {Lamb}, D.~Q., {Miknaitis}, G., {Munn}, J.~A.,
  {Nichol}, R., {Okamura}, S., {Pier}, J.~R., {Prada}, F., {Richards}, G.~T.,
  {Szalay}, A., \& {York}, D.~G. 2001, \aj, 122, 2833

\bibitem[{{Fan} {et~al.}(2006{\natexlab{b}}){Fan}, {Strauss}, {Becker},
  {White}, {Gunn}, {Knapp}, {Richards}, {Schneider}, {Brinkmann}, \&
  {Fukugita}}]{2006Fan}
{Fan}, X., {Strauss}, M.~A., {Becker}, R.~H., {White}, R.~L., {Gunn}, J.~E.,
  {Knapp}, G.~R., {Richards}, G.~T., {Schneider}, D.~P., {Brinkmann}, J., \&
  {Fukugita}, M. 2006{\natexlab{b}}, \aj, 132, 117

\bibitem[{{Furlanetto} {et~al.}(2004){Furlanetto}, {Zaldarriaga}, \&
  {Hernquist}}]{2004FZH}
{Furlanetto}, S.~R., {Zaldarriaga}, M., \& {Hernquist}, L. 2004, \apj, 613, 1

\bibitem[{{Gao} {et~al.}(2005){Gao}, {White}, {Jenkins}, {Frenk}, \&
  {Springel}}]{2005Gao}
{Gao}, L., {White}, S.~D.~M., {Jenkins}, A., {Frenk}, C.~S., \& {Springel}, V.
  2005, \mnras, 363, 379

\bibitem[{{Gnedin}(2000)}]{2000Gnedin}
{Gnedin}, N.~Y. 2000, \apj, 535, 530

\bibitem[{{Gnedin} \& {Fan}(2006)}]{2006GnedinFan}
{Gnedin}, N.~Y., \& {Fan}, X. 2006, \apj, 648, 1

\bibitem[{{G{\'o}rski} {et~al.}(2002){G{\'o}rski}, {Banday}, {Hivon}, \&
  {Wandelt}}]{2002Healpix}
{G{\'o}rski}, K.~M., {Banday}, A.~J., {Hivon}, E., \& {Wandelt}, B.~D. 2002, in
  ASP Conf. Ser. 281: Astronomical Data Analysis Software and Systems XI, ed.
  D.~A. {Bohlender}, D.~{Durand}, \& T.~H. {Handley}, 107--+

\bibitem[{{G{\'o}rski} {et~al.}(2005){G{\'o}rski}, {Hivon}, {Banday},
  {Wandelt}, {Hansen}, {Reinecke}, \& {Bartelmann}}]{2005Healpix}
{G{\'o}rski}, K.~M., {Hivon}, E., {Banday}, A.~J., {Wandelt}, B.~D., {Hansen},
  F.~K., {Reinecke}, M., \& {Bartelmann}, M. 2005, \apj, 622, 759

\bibitem[{{Gunn} \& {Peterson}(1965)}]{1965GunnPeterson}
{Gunn}, J.~E., \& {Peterson}, B.~A. 1965, \apj, 142, 1633

\bibitem[{{Haiman} {et~al.}(1997){Haiman}, {Rees}, \& {Loeb}}]{1997HRL}
{Haiman}, Z., {Rees}, M.~J., \& {Loeb}, A. 1997, \apj, 476, 458

\bibitem[{{Hernquist} \& {Springel}(2003)}]{2003HernquistSpringel}
{Hernquist}, L., \& {Springel}, V. 2003, \mnras, 341, 1253

\bibitem[{{Iliev} {et~al.}(2006{\natexlab{a}}){Iliev}, {Mellema}, {Pen},
  {Merz}, {Shapiro}, \& {Alvarez}}]{2006Iliev}
{Iliev}, I.~T., {Mellema}, G., {Pen}, U.-L., {Merz}, H., {Shapiro}, P.~R., \&
  {Alvarez}, M.~A. 2006{\natexlab{a}}, \mnras, 369, 1625

\bibitem[{{Iliev} {et~al.}(2006{\natexlab{b}}){Iliev}, {Mellema}, {Shapiro}, \&
  {Pen}}]{2006IMSP}
{Iliev}, I.~T., {Mellema}, G., {Shapiro}, P.~R., \& {Pen}, U.-L.
  2006{\natexlab{b}}, ArXiv Astrophysics e-prints

\bibitem[{{Kashikawa} {et~al.}(2006){Kashikawa}, {Shimasaku}, {Malkan}, {Doi},
  {Matsuda}, {Ouchi}, {Taniguchi}, {Ly}, {Nagao}, {Iye}, {Motohara},
  {Murayama}, {Murozono}, {Nariai}, {Ohta}, {Okamura}, {Sasaki}, {Shioya}, \&
  {Umemura}}]{2006Kashikawa}
{Kashikawa}, N., {Shimasaku}, K., {Malkan}, M.~A., {Doi}, M., {Matsuda}, Y.,
  {Ouchi}, M., {Taniguchi}, Y., {Ly}, C., {Nagao}, T., {Iye}, M., {Motohara},
  K., {Murayama}, T., {Murozono}, K., {Nariai}, K., {Ohta}, K., {Okamura}, S.,
  {Sasaki}, T., {Shioya}, Y., \& {Umemura}, M. 2006, \apj, 648, 7

\bibitem[{{Kogut} {et~al.}(2003){Kogut}, {Spergel}, {Barnes}, {Bennett},
  {Halpern}, {Hinshaw}, {Jarosik}, \& {Limon}}]{2003KogutWMAP}
{Kogut}, A., {Spergel}, D.~N., {Barnes}, C., {Bennett}, C.~L., {Halpern}, M.,
  {Hinshaw}, G., {Jarosik}, N., \& {Limon}, M. 2003, \apjs, 148, 161

\bibitem[{{Kohler} {et~al.}(2005){Kohler}, {Gnedin}, \& {Hamilton}}]{2005KGH}
{Kohler}, K., {Gnedin}, N.~Y., \& {Hamilton}, A.~J.~S. 2005, ArXiv Astrophysics
  e-prints

\bibitem[{{Komatsu} \& {Seljak}(2001)}]{2001KomatsuSeljak}
{Komatsu}, E., \& {Seljak}, U. 2001, \mnras, 327, 1353

\bibitem[{{Lacey} \& {Cole}(1993)}]{1993LaceyCole}
{Lacey}, C., \& {Cole}, S. 1993, \mnras, 262, 627

\bibitem[{{Lacey} \& {Cole}(1994)}]{1994LaceyCole}
---. 1994, \mnras, 271, 676

\bibitem[{{Lukic} {et~al.}(2007){Lukic}, {Heitmann}, {Habib}, {Bashinsky}, \&
  {Ricker}}]{2007LukicHHBR}
{Lukic}, Z., {Heitmann}, K., {Habib}, S., {Bashinsky}, S., \& {Ricker}, P.~M.
  2007, ArXiv Astrophysics e-prints

\bibitem[{{Madau} {et~al.}(2004){Madau}, {Rees}, {Volonteri}, {Haardt}, \&
  {Oh}}]{2004Madau}
{Madau}, P., {Rees}, M.~J., {Volonteri}, M., {Haardt}, F., \& {Oh}, S.~P. 2004,
  \apj, 604, 484

\bibitem[{{McQuinn} {et~al.}(2007){McQuinn}, {Lidz}, {Zahn}, {Dutta},
  {Hernquist}, \& {Zaldarriaga}}]{2007McQuinn}
{McQuinn}, M., {Lidz}, A., {Zahn}, O., {Dutta}, S., {Hernquist}, L., \&
  {Zaldarriaga}, M. 2007, \mnras, 377, 1043

\bibitem[{{Mesinger} \& {Furlanetto}(2007)}]{2007Mesinger}
{Mesinger}, A., \& {Furlanetto}, S. 2007, ArXiv e-prints, 704

\bibitem[{{Miralda-Escud{\'e}} {et~al.}(2000){Miralda-Escud{\'e}}, {Haehnelt},
  \& {Rees}}]{2000MiraldaEscude}
{Miralda-Escud{\'e}}, J., {Haehnelt}, M., \& {Rees}, M.~J. 2000, \apj, 530, 1

\bibitem[{{Mo} \& {White}(1996)}]{1996MoWhite}
{Mo}, H.~J., \& {White}, S.~D.~M. 1996, \mnras, 282, 347

\bibitem[{{Nagamine} {et~al.}(2004){Nagamine}, {Cen}, {Hernquist}, {Ostriker},
  \& {Springel}}]{2004NCHOS}
{Nagamine}, K., {Cen}, R., {Hernquist}, L., {Ostriker}, J.~P., \& {Springel},
  V. 2004, \apj, 610, 45

\bibitem[{{Navarro} {et~al.}(1997){Navarro}, {Frenk}, \& {White}}]{1997NFW}
{Navarro}, J.~F., {Frenk}, C.~S., \& {White}, S.~D.~M. 1997, \apj, 490, 493

\bibitem[{{O'Shea} \& {Norman}(2006)}]{2006OsheaNorman}
{O'Shea}, B.~W., \& {Norman}, M.~L. 2006, \apj, 648, 31

\bibitem[{{Ostriker} {et~al.}(2005){Ostriker}, {Bode}, \&
  {Babul}}]{2005OstrikerBB}
{Ostriker}, J.~P., {Bode}, P., \& {Babul}, A. 2005, \apj, 634, 964

\bibitem[{{Page} {et~al.}(2006){Page}, {Hinshaw}, {Komatsu}, {Nolta},
  {Spergel}, {Bennett}, {Barnes}, {Bean}, {Dore'}, {Halpern}, {Hill},
  {Jarosik}, {Kogut}, {Limon}, {Meyer}, {Odegard}, {Peiris}, {Tucker}, {Verde},
  {Weiland}, {Wollack}, \& {Wright}}]{2006PageWMAP3}
{Page}, L., {Hinshaw}, G., {Komatsu}, E., {Nolta}, M.~R., {Spergel}, D.~N.,
  {Bennett}, C.~L., {Barnes}, C., {Bean}, R., {Dore'}, O., {Halpern}, M.,
  {Hill}, R.~S., {Jarosik}, N., {Kogut}, A., {Limon}, M., {Meyer}, S.~S.,
  {Odegard}, N., {Peiris}, H.~V., {Tucker}, G.~S., {Verde}, L., {Weiland},
  J.~L., {Wollack}, E., \& {Wright}, E.~L. 2006, ArXiv Astrophysics e-prints

\bibitem[{{Paschos} \& {Norman}(2005)}]{2005PaschosNorman}
{Paschos}, P., \& {Norman}, M.~L. 2005, \apj, 631, 59

\bibitem[{{Peebles}(1980)}]{1980Peebles}
{Peebles}, P.~J.~E. 1980, {The large-scale structure of the universe} (Research
  supported by the National Science Foundation.~Princeton, N.J., Princeton
  University Press, 1980.~435 p.)

\bibitem[{{Press} \& {Schechter}(1974)}]{1974PressSchechter}
{Press}, W.~H., \& {Schechter}, P. 1974, \apj, 187, 425

\bibitem[{{Razoumov} {et~al.}(2002){Razoumov}, {Norman}, {Abel}, \&
  {Scott}}]{2002Razoumov}
{Razoumov}, A.~O., {Norman}, M.~L., {Abel}, T., \& {Scott}, D. 2002, \apj, 572,
  695

\bibitem[{{Reed} {et~al.}(2007){Reed}, {Bower}, {Frenk}, {Jenkins}, \&
  {Theuns}}]{2007ReedBFJT}
{Reed}, D.~S., {Bower}, R., {Frenk}, C.~S., {Jenkins}, A., \& {Theuns}, T.
  2007, \mnras, 374, 2

\bibitem[{{Ricotti} \& {Ostriker}(2004)}]{2004RicottiOstriker}
{Ricotti}, M., \& {Ostriker}, J.~P. 2004, \mnras, 350, 539

\bibitem[{{Schaerer}(2002)}]{2002Schaerer}
{Schaerer}, D. 2002, \aap, 382, 28

\bibitem[{{Schaerer}(2003)}]{2003Schaerer}
---. 2003, \aap, 397, 527

\bibitem[{{Seljak} \& {Zaldarriaga}(1996)}]{1996SeljakZaldarriagaCMBFAST}
{Seljak}, U., \& {Zaldarriaga}, M. 1996, \apj, 469, 437

\bibitem[{{Sheth} \& {Tormen}(1999)}]{1999ShethTormen}
{Sheth}, R.~K., \& {Tormen}, G. 1999, \mnras, 308, 119

\bibitem[{{Shin} {et~al.}(2007){Shin}, {Trac}, \& {Cen}}]{2007Shin}
{Shin}, M.-S., {Trac}, H., \& {Cen}, R. 2007, ArXiv e-prints, 708

\bibitem[{{Sokasian} {et~al.}(2003){Sokasian}, {Abel}, {Hernquist}, \&
  {Springel}}]{2003Sokasian}
{Sokasian}, A., {Abel}, T., {Hernquist}, L., \& {Springel}, V. 2003, \mnras,
  344, 607

\bibitem[{{Sokasian} {et~al.}(2001){Sokasian}, {Abel}, \&
  {Hernquist}}]{2001Sokasian}
{Sokasian}, A., {Abel}, T., \& {Hernquist}, L.~E. 2001, New Astronomy, 6, 359

\bibitem[{{Spergel} {et~al.}(2007){Spergel}, {Bean}, {Dor{\'e}}, {Nolta},
  {Bennett}, {Dunkley}, {Hinshaw}, {Jarosik}, {Komatsu}, {Page}, {Peiris},
  {Verde}, {Halpern}, {Hill}, {Kogut}, {Limon}, {Meyer}, {Odegard}, {Tucker},
  {Weiland}, {Wollack}, \& {Wright}}]{2007SpergelWMAP3}
{Spergel}, D.~N., {Bean}, R., {Dor{\'e}}, O., {Nolta}, M.~R., {Bennett}, C.~L.,
  {Dunkley}, J., {Hinshaw}, G., {Jarosik}, N., {Komatsu}, E., {Page}, L.,
  {Peiris}, H.~V., {Verde}, L., {Halpern}, M., {Hill}, R.~S., {Kogut}, A.,
  {Limon}, M., {Meyer}, S.~S., {Odegard}, N., {Tucker}, G.~S., {Weiland},
  J.~L., {Wollack}, E., \& {Wright}, E.~L. 2007, \apjs, 170, 377

\bibitem[{{Springel} \& {Hernquist}(2003)}]{2003SpringelHernquist}
{Springel}, V., \& {Hernquist}, L. 2003, \mnras, 339, 289

\bibitem[{{Tasitsiomi}(2006)}]{2006Tasitsiomi}
{Tasitsiomi}, A. 2006, \apj, 645, 792

\bibitem[{{Theuns} {et~al.}(1998){Theuns}, {Leonard}, {Efstathiou}, {Pearce},
  \& {Thomas}}]{1998Theuns}
{Theuns}, T., {Leonard}, A., {Efstathiou}, G., {Pearce}, F.~R., \& {Thomas},
  P.~A. 1998, \mnras, 301, 478

\bibitem[{{Trac} \& {Pen}(2006)}]{2006TracPenOCH}
{Trac}, H., \& {Pen}, U.-L. 2006, New Astronomy, 11, 273

\bibitem[{{Warren} {et~al.}(2006){Warren}, {Abazajian}, {Holz}, \&
  {Teodoro}}]{2006WarrenAHT}
{Warren}, M.~S., {Abazajian}, K., {Holz}, D.~E., \& {Teodoro}, L. 2006, \apj,
  646, 881

\bibitem[{{Wyithe} \& {Loeb}(2003)}]{2003WyitheLoeb}
{Wyithe}, J.~S.~B., \& {Loeb}, A. 2003, \apj, 586, 693

\bibitem[{{Wyithe} \& {Cen}(2006)}]{2006WyitheCen}
{Wyithe}, S., \& {Cen}, R. 2006, ArXiv Astrophysics e-prints

\bibitem[{{Yoshida} {et~al.}(2004){Yoshida}, {Bromm}, \& {Hernquist}}]{2004YBH}
{Yoshida}, N., {Bromm}, V., \& {Hernquist}, L. 2004, \apj, 605, 579

\bibitem[{{Yoshida} {et~al.}(2006){Yoshida}, {Omukai}, {Hernquist}, \&
  {Abel}}]{2006YOHA}
{Yoshida}, N., {Omukai}, K., {Hernquist}, L., \& {Abel}, T. 2006, \apj, 652, 6

\bibitem[{{Zahn} {et~al.}(2007){Zahn}, {Lidz}, {McQuinn}, {Dutta}, {Hernquist},
  {Zaldarriaga}, \& {Furlanetto}}]{2007Zahn}
{Zahn}, O., {Lidz}, A., {McQuinn}, M., {Dutta}, S., {Hernquist}, L.,
  {Zaldarriaga}, M., \& {Furlanetto}, S.~R. 2007, \apj, 654, 12

\end{thebibliography}

\end{document}